\documentclass[letterpaper, 11pt, conference, onecolumn]{ieeeconf}

\IEEEoverridecommandlockouts                              
\usepackage{epsfig}
\usepackage{url}
\usepackage{multicol}
\usepackage{amssymb}
\usepackage[tbtags]{amsmath}
\usepackage{latexsym}
\usepackage{euscript}
\usepackage{graphics}

\usepackage[tbtags]{amsmath} 
\usepackage{amssymb}  
\usepackage{verbatim} 

\DeclareMathOperator*{\argmax}{arg\,max}

\newcounter{actr}
{\begin{list}{(\alph{actr})}{\usecounter{actr}}}{\end{list}}

\newcounter{ictr}
{\begin{list}{(\roman{ictr})}{\usecounter{ictr}}}{\end{list}}

\newtheorem{theorem}{Theorem}

 \newtheorem{corollary}{Corollary}
 \newtheorem{proposition}{Proposition}

\newcommand{\qed}{\rule[0.1ex]{1.4ex}{1.6ex}}

\newcounter{psctr}
\newcounter{probctr}[psctr]

\DeclareMathAlphabet{\mathbsf}{OT1}{cmss}{bx}{n}
\DeclareMathAlphabet{\mathssf}{OT1}{cmss}{m}{sl}

\DeclareSymbolFont{bsfletters}{OT1}{cmss}{bx}{n}
\DeclareSymbolFont{ssfletters}{OT1}{cmss}{m}{n}
\DeclareMathSymbol{\bsfGamma}{0}{bsfletters}{'000}
\DeclareMathSymbol{\ssfGamma}{0}{ssfletters}{'000}
\DeclareMathSymbol{\bsfDelta}{0}{bsfletters}{'001}
\DeclareMathSymbol{\ssfDelta}{0}{ssfletters}{'001}
\DeclareMathSymbol{\bsfTheta}{0}{bsfletters}{'002}
\DeclareMathSymbol{\ssfTheta}{0}{ssfletters}{'002}
\DeclareMathSymbol{\bsfLambda}{0}{bsfletters}{'003}
\DeclareMathSymbol{\ssfLambda}{0}{ssfletters}{'003}
\DeclareMathSymbol{\bsfXi}{0}{bsfletters}{'004}
\DeclareMathSymbol{\ssfXi}{0}{ssfletters}{'004}
\DeclareMathSymbol{\bsfPi}{0}{bsfletters}{'005}
\DeclareMathSymbol{\ssfPi}{0}{ssfletters}{'005}
\DeclareMathSymbol{\bsfSigma}{0}{bsfletters}{'006}
\DeclareMathSymbol{\ssfSigma}{0}{ssfletters}{'006}
\DeclareMathSymbol{\bsfUpsilon}{0}{bsfletters}{'007}
\DeclareMathSymbol{\ssfUpsilon}{0}{ssfletters}{'007}
\DeclareMathSymbol{\bsfPhi}{0}{bsfletters}{'010}
\DeclareMathSymbol{\ssfPhi}{0}{ssfletters}{'010}
\DeclareMathSymbol{\bsfPsi}{0}{bsfletters}{'011}
\DeclareMathSymbol{\ssfPsi}{0}{ssfletters}{'011}
\DeclareMathSymbol{\bsfOmega}{0}{bsfletters}{'012}
\DeclareMathSymbol{\ssfOmega}{0}{ssfletters}{'012}

\newcommand{\rva}{{\mathssf{a}}}    

\newcommand{\sva}{a}
\newcommand{\svb}{b}

\newcommand{\rvb}{{\mathssf{b}}}    
\newcommand{\rvc}{{\mathssf{c}}}    

\newcommand{\svx}{x}            

\newcommand{\rvy}{{\mathssf{y}}}    

\newcommand{\svy}{y}

 \usepackage{fullpage}

\begin{document}

\title{\LARGE \bf  Joint source-channel with side information coding error exponents  }
\author{Cheng Chang
\thanks{Cheng Chang is with Hewlett-Packard Labs, Palo Alto.
        {\tt\small Email:  cchang@eecs.berkeley.edu}}%
}

\maketitle

\begin{abstract}In this paper, we    study
the upper and the lower bounds on the joint source-channel coding
error exponent with decoder side-information. The results in the
paper  are  non-trivial extensions of the Csisz\'{a}r's classical
paper~\cite{Csiszar:80}. Unlike the joint source-channel coding
result in~\cite{Csiszar:80}, it is not obvious whether the lower
bound and the upper bound are equivalent even if the channel coding
error exponent is known. For a class of channels, including the
symmetric channels, we apply  a game-theoretic result to establish
the existence of a  saddle point and hence prove that the lower and
upper bounds are the same if the channel coding error exponent is
known. More interestingly, we show that encoder side-information
does not increase the error exponents in this case.
\end{abstract}

\section{Introduction}
\begin{figure}[htbp]
\setlength{\unitlength}{0.6mm}
\begin{picture}(100,45)(-60,0)

\put(40,30){\line(1,0){26}} \put(40,40){\line(1,0){26}}
\put(40,30){\line(0,1){10}} \put(66, 30){\line(0,1){10}}
\put(42,34){Encoder }

\put(80,30){\line(1,0){36}} \put(80,40){\line(1,0){36}}
\put(80,30){\line(0,1){10}} \put(116, 30){\line(0,1){10}}
\put(82,34){DMC $W_{Y|X}$}

\put(126,30){\line(1,0){26}} \put(126,40){\line(1,0){26}}
\put(126,30){\line(0,1){10}} \put(152, 30){\line(0,1){10}}
\put(128,34){Decoder }

\put(116, 34){\vector(1,0){10}} \put(66, 34){\vector(1,0){12}}
\put(152, 34){\vector(1,0){10}}

 \put(30, 5){\line(1,0){110}}
\put(140, 5){\vector(0,1){25}}

\put(162,34){ $\widehat{\rva}^n$}

\put(30, 35){\vector(1,0){10}}

\put(10, 34){ ${\rva}^n$}\put(10, 2){ ${\rvb}^n$}

\put(0, 20){ $(\rva_i,\rvb_i)\sim P_{AB}$}

\put(13,24){\vector(0,1){8}} \put(13,16){\vector(0,-1){8}}
 \end{picture}
     \caption[]{Source coding with decoder  side-information}
     \label{fig.setup}
 \end{figure}
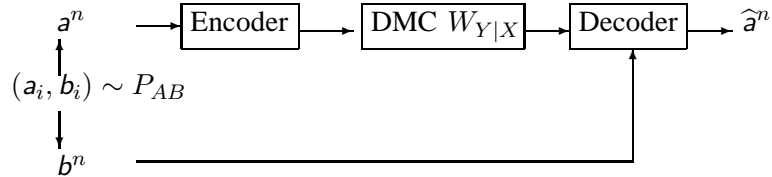

 In Shannon's very first paper on information
theory~\cite{Shannon}, it is established that separate coding is
optimal for memoryless source channel pairs. Reliable communication
is possible if and only if the entropy of the source is lower than
the capacity of the channel.  However, the story is different when
error exponent is considered. It is shown that joint source-channel
coding achieves strictly better error exponent than
separate\footnote{In~\cite{Csiszar:80}, Csisz\'{a}r hand-wavily
shows that the obvious separate coding scheme is suboptimal in terms
achieving the best error exponent.  The rather obvious result is
rigidly proved in~\cite{Zhong_phd}. } coding~\cite{Csiszar:80}. The
key technical component of~\cite{Csiszar:80} is a channel coding
scheme to protect different message sets with different channel
coding error exponents. In this paper, we are concerned with the
joint source-channel coding with side information problem as shown
in Figure~\ref{fig.setup}. For a special setup of
Figure~\ref{fig.setup}, where the discrete memoryless channel (DMC)
is a noiseless channel with capacity\footnote{In this paper, we use
bits and $\log_2$, and $R$ is always non-negative. } $R$, i.e. the
source coding with side-information problem,  the reliable
reconstruction of $\rva^n$ at the decoder is possible if and only if
$R$ is larger than the conditional entropy
$H(P_{A|B})$~\cite{Slepian_Wolf}. The error exponents of this
problem is also studied in~\cite{Gallager_Report_76,Csiszar} and
more importantly in~\cite{Ahlswede_dual}.

The duality between source coding with decoder side-information and
channel coding is established in the 80's~\cite{Ahlswede_dual}. This
is an important result that all the channel coding error exponent
bounds can be easily applied to source coding with side-information
error exponent.    The result is a consequence of the type covering
lemma~\cite{Csiszar}, also known as the Johnson-Stein-Lov\'{a}sz
theorem~\cite{Cohen_covering_code}. With this duality result, we
know that the error exponent of channel coding of channel $W_{Y|X}$
with channel code composition $Q_X$ is essentially the same problem
as the error exponent of source coding with decoder side-information
where the joint distribution is $Q_X\times W_{Y|X}$. Hence a natural
question is what if we put these two dual problems together, what is
the error exponent of joint source-channel coding with decoder
side-information?

The more general case, where $W_{Y|X}$ is a noisy channel, is
recently studied~\cite{Zhong_Alajaji,Zhong_phd}. It is shown that,
not surprisingly, the reliable reconstruction of $\rva^n$ is
possible if and only if the channel capacity of the channel is
larger than the conditional entropy of the source.  A suboptimal
error exponent based on a mixture scheme of  separate coding and the
joint source channel coding first developed in~\cite{Csiszar:80} is
achieved. In this paper, we follow Csisz\'{a}r's idea
in~\cite{Csiszar:80} and develop a new coding scheme for joint
source channel coding with decoder side-information. For a class of
channels, including the symmetric channels, the resulted lower and
upper bounds have the same property as the joint source-channel
coding  error exponent \textit{without}  side-information
in~\cite{Csiszar:80}: they match if the channel coding error
exponent is known at a critical rate. We use a game theoretic
approach to interpret  this result.

The outline of the paper is as follows. We review the problem setup
and classical error exponent results in Section~\ref{sec.intro}.
Then in Section~\ref{sec.both}, we present the error exponent result
for joint source-channel coding with both decoder and encoder side
information which provides a simple upper bound to the error
exponent investigated in the paper. This is a simple corollary of
Theorem 5 in~\cite{Csiszar:80}. The main result of this paper is
presented in Section~\ref{sec.mainresult}.  Some implications of
these bounds are given in Section~\ref{sec.discussion}.

 \section{Review of source and channel doing error exponents }\label{sec.intro}
In this paper random variables are denoted by $\rva$ and $\rvb$, the
realizations of the random variables are denoted by $\sva$ and
$\svb$.

\subsection{System model of joint source-channel coding with decoder side-information }

As shown in Figure~\ref{fig.setup}, the source and side-information,
$\rva^n$ and $ \rvb^n$ respectively, are random variables i.i.d from
distribution $P_{AB}$ on a finite alphabet $\mathcal A\times
\mathcal B$. The channel is memoryless with input/output probability
transition $W_{Y|X}$, where the input/output alphabets $\mathcal X$
and $\mathcal Y$ are finite. Without loss of generality, we assume
that the number of source symbols and the number of channel uses are
equal, i.e. the encoder observes $\sva^n$ and sends a codeword
$x^n(\sva^n)$ of length $n$ to the channel, the decoder observes the
channel output $\svy^n$ and side-information $\svb^n$ which is not
available to the encoder, the estimate is $\widehat\sva^n(\svb^n,
\svy^n)$.

The error probability is the expectation of the decoding error
average over all channel  and source behaviors.
\begin{eqnarray}
\Pr(\rva^n \neq \widehat \rva^n(\rvb^n, \rvy^n))=\sum_{a^n, b^n}
P_{AB}(a^n, b^n)\sum_{y^n} W_{Y|X}(y^n| x^n(a^n)) 1(a^n \neq
\widehat a^n (b^n, y^n))\label{eqn.errorP_def}.
\end{eqnarray}
The error exponent, for  the optimal coding scheme, is defined as
\begin{eqnarray}
   E(P_{AB},
W_{Y|X})=\lim\inf_{n\rightarrow \infty}-\frac{1}{n} \log \Pr(\rva^n
\neq \widehat \rva^n(\rvb^n, \rvy^n)).\label{eqn.EE_def}
\end{eqnarray}
  The main result of this paper is to establish both upper and lower
  bounds on $E(P_{AB},
W_{Y|X})$ and show the tightness of these bounds.

\subsection{Classical error exponent results}\footnote{
In this paper, we write the error exponents (both channel coding and
source coding) in the style of Csisz\'{a}r's  method of types,
equivalent Gallager style error exponents can be derived through the
Fenchel duality.} We review some classical results on channel coding
error exponents and source coding with side-information error
exponents. These bounds are investigated
in~\cite{Gallager},~\cite{Csiszar},~\cite{Gallager_Report_76} and
~\cite{Gallager_sphere}.

\subsubsection{Channel coding error exponents $E_c(R,W_{Y|X})$}
Channel coding is a special case of  joint source-channel coding
with side-information: the source $\rva$ and the side-information
$\rvb$ are independent, i.e. $P_{AB}=P_A\times P_B$, and $\rva$ is a
uniform distributed random variable on   $\{1,2,...,2^R\}$. For the
sake of simplicity, we assume that $2^R$ is an integer. This is not
a problem if $2^R$ is not an integer since we can lump $K$ symbols
together and approximate $2^{KR}$ by an integer for some $K$, this
is not a problem because $\lim\limits_{K\rightarrow
\infty}\frac{1}{K}\log_2(\lfloor 2^{KR} \rfloor)=R$. With this
interpretation of channel coding, the definitions of error
probability in~(\ref{eqn.errorP_def}) and error exponent
in~(\ref{eqn.EE_def}) still holds.

The channel coding error exponent $E_c(R,W_{Y|X})$ is lower bounded
by the random coding error exponent  and upper bounded by the sphere
packing error exponent.
\begin{eqnarray}
E_{r}(R, W_{Y|X})\leq E_c(R,W_{Y|X})\leq
E_{sp}(R,W_{Y|X})\label{eqn.erandom}
\end{eqnarray}
\begin{eqnarray}
\mbox{where } E_{r}(R,
W_{Y|X})&=&\max_{S_X}\inf_{V_{Y|X}}D(V_{Y|X}\|W_{Y|X} | S_X)+
|I(V_{Y|X};S_X)-R|^+
 \label{eqn.channel_random_ee}\\
 &=&\max_{S_X}E_{r}(R,S_X,W_{Y|X})\nonumber\\
 \mbox{and }
E_{sp}(R,W_{Y|X}) &=&
\max_{S_X}\inf_{V_{Y|X}:I(V_{Y|X};S_X)<R}D(V_{Y|X}\|W_{Y|X} |
S_X)\label{eqn.channel_sphere_ee}\\
&=&\max_{S_X}E_{sp}(R,S_X,W_{Y|X}) \nonumber
\end{eqnarray}

Here $S_X$ is the input composition (type) of the code words.
$E_{r}(R,W_{Y|X})=E_{sp}(R,W_{Y|X})$ in the high rate regime that
$R>R_{cr}$ where $R_{cr}$ is defined in~\cite{Gallager} as   the
minimum rate for which the sphere packing $E_{sp}(R,W_{Y|X})$ and
random coding error exponents $E_{r}(R,W_{Y|X})$ match for channel
$W_{Y|X}$.   There are tighter bounds on the channel coding error
exponents $E_{c}(R, W_{Y|X})$ in the low rate regime for $R<R_{cr}$,
known as straight-line lower bounds and expurgation upper
bounds~\cite{Gallager}.  However, in this paper, we focus on the
basic random coding and sphere packing bounds, as the main message
can be effectively carried out.

It is well known~\cite{Gallager} that both the random coding and the
sphere-packing bounds are decreasing with $R$ and are convex in $R$.
And they are both positive if and only if $R< C(W_{Y|X})$, where
$C(W_{Y|X})$ is the capacity of the channel $W_{Y|X}$.
\vspace{0.1in}

\subsubsection{Source coding with decoder side-information error
exponents} This is also a special case of the general setup in
Figure~\ref{fig.setup}.  This time the channel $W_{Y|X}$ is a
noiseless channel with input-output alphabet $\mathcal X=\mathcal Y$
and $|\mathcal X|=2^{R}$. Again, we can reasonably assume that
$2^{R}$ is an integer.

The source coding with side-information error exponent\footnote{ In
this paper, if  $R\geq \log_2|\mathcal A|$ for source coding with
side-information error exponents, we let the error exponent be
$\infty$.} $e(R, P_{AB})$ can be bounded as follows:
\begin{eqnarray}
e_{L}(R, P_{AB})\leq e(R,P_{AB})\leq e_{U}(R,P_{AB})
\end{eqnarray}
\begin{eqnarray}
\mbox{where }&& e_{L}(R, P_{AB})= \inf_{Q_{AB}}D(Q_{AB}\|P_{AB})+ |R-H(Q_{A|B})|^+ \nonumber\\
 && e_{U}(R, P_{AB})= \inf_{Q_{AB}: H(Q_{A|B})>R}D(Q_{AB}\|P_{AB}). \nonumber
\end{eqnarray}

The duality between channel coding and source coding with decoder
side information had been well understood~\cite{Ahlswede_dual}. We
give the following duality results on error exponents. .
\begin{eqnarray}
e(R, Q_A, P_{B|A})&=&E_c(H(Q_A)-R, Q_A, P_{B|A})\nonumber\\
\mbox{or equivalently : } e(H(Q_A)-R, Q_A, P_{B|A})&=&E_c(R, Q_A,
P_{B|A})\nonumber
\end{eqnarray}
where $E_c(R, Q_A, P_{B|A})$ is the channel coding error exponent
for channel $P_{B|A}$ at rate $R$ and the codebook composition is
$Q_A$. $e(R, Q_A, P_{B|A})$ is the source coding with side
information error exponent at rate $R$ with source sequences
uniformly distributed in type $Q_A$ and the side information is the
output of channel $P_{B|A}$ with input sequence of type $Q_A$. So
obviously, we have:
\begin{eqnarray}
&& E_c(R,  P_{B|A})=\max_{Q_A}\{ E_c(R, Q_A, P_{B|A})\}\nonumber \\
&&e(R, P_{AB})=\min_{Q_A} \{D(Q_A\|P_A)+e(R, Q_A,
P_{B|A})\}\nonumber
\end{eqnarray}

 These
results are established by the type covering lemma~\cite{Csiszar:80}
on the operational level, i.e. a complete characterizations of the
source coding with side information error exponent $e(R, Q_A,
P_{B|A}) $ implies a complete characterizations of the channel
coding error exponent $E_c(H(Q_A)-R, Q_A, P_{B|A})$ and vice versa.

From these duality results, it is well known that both the lower and
the upper bounds are increasing with $R$ and are convex in $R$. And
they are both positive if and only if $R> H(P_{A|B})$. The special
case of the source coding with decoder side information problem is
that the side information is independent of the source, i.e.
$P_{AB}=P_A\times P_B$. In this case, the error exponent is
completely characterized~\cite{Csiszar},
\begin{eqnarray}
e(R, P_A)=\inf_{Q_{A}: H(Q_{A})>R}D(Q_{A}\|P_{A})
\end{eqnarray}

\vspace{0.1in}

\subsubsection{Joint source-channel coding error
exponents~\cite{Csiszar:80}}\label{sec.jsc} In his seminal
paper~\cite{Csiszar:80}, the joint source-channel coding error
exponents is studied. This is yet another special case of the
general setup in Figure~\ref{fig.setup}. When $\rva$ and $\rvb$ are
independent, i.e. $P_{AB}=P_A\times P_B$, we can drop all the $b$
terms in~(\ref{eqn.errorP_def}). Hence the error probability is
defined as:
\begin{eqnarray}
\Pr(\rva^n \neq \widehat \rva^n( \rvy^n))=\sum_{a^n}
P_{A}(a^n)\sum_{y^n} W_{Y|X}(y^n| x^n(a^n)) 1(a^n \neq \widehat a^n
(y^n))\label{eqn.errorPAonly_def}.
\end{eqnarray}
Write the error exponent of~(\ref{eqn.errorPAonly_def}) as $E(P_A,
W_{Y|X})$. The lower and upper bounds of the error exponents are
derived in~\cite{Csiszar:80}. It is shown that:
\begin{eqnarray}
 \min_{R}  \{e(R, P_A) +E_{sp}(R,W_{Y|X})\} \leq E(P_A,
W_{Y|X}) \leq \min_{R}  \{e(R, P_A)
+E_{r}(R,W_{Y|X})\}\label{eqn.jsc_csiszar}
\end{eqnarray}
The upper bound is derived by using standard method of types
argument. The lower bound is a direct consequence of the channel
coding Theorem~5 in~\cite{Csiszar:80}.

The difference between the lower and upper bounds is in the channel
coding error exponent.  The joint source channel coding error
exponent is ``almost'' completely characterized because the only
possible improvement is to determine the channel coding error
exponent which is still not completely characterized in the low rate
regime where $R< R_{cr}$. However, let  $R^*$ be the rate that
minimizes $\{e(R, P_A) +E_{r}(R,W_{Y|X})\}$, if $R^*\geq R_{cr}$ or
equivalently $E_{r}(R^*,W_{Y|X})=E_{sp}(R^*,W_{Y|X})$, then we have
a complete characterization of the joint source channel coding error
exponent:
\begin{eqnarray}
 E(P_A, W_{Y|X})=e(R^*, P_A)+E_{r}(R^*,W_{Y|X})\label{eqn.jsc_complete}.
\end{eqnarray}
The goal of this paper is to derive a similar result for $ E(P_{AB},
W_{Y|X})$ defined in~(\ref{eqn.EE_def}) as that for the joint source
channel coding in~(\ref{eqn.jsc_csiszar})
and~(\ref{eqn.jsc_complete}).

\vspace{0.1in}

\subsubsection{A recite of Theorem 5 in ~\cite{Csiszar:80}} Given a
sequence of positive
 integers $\{m_n\}$ with $\frac{1}{n}\log m_n \rightarrow 0$ and
 $m_n$ message sets $\mathcal A_1,.... \mathcal A_{m_n}$ each with
 size $|\mathcal A_i|=2^{nR_i}$. Then there exists a channel code $(f_0,
 \phi_0)$, where the encoder $f_0: \bigcup_{i=0}^{m_n} \mathcal A_i\rightarrow \mathcal
 X^n$ where $f_0(a)=x^n(a)\in S_X^i$ for $a\in \mathcal A_i$ and the decoder $\phi_0: \mathcal Y^n\rightarrow \bigcup_{i=0}^{m_n} \mathcal
 A_i$, write $ \phi_0(y^n)$ as $\widehat a(y^n)$ s.t.  for any message $a\in \mathcal
 A_i$, the decoding error
 $$ p_e(a)= \sum_{y^n} W_{Y|X}(y^n| x^n(a)) 1(a \neq
\widehat a ( y^n))\leq
 2^{n(E_{r}(R_i,S^i_X,W_{Y|X})-\epsilon_n)}$$
for every channel $W_{Y|X}$, and $\epsilon_n\rightarrow 0$. In
particular, if the channel $W_{Y|X}$ is known to the encoder, each
$S^i_X$ can be picked to maximize $E_{r}(R_i,S^i_X,W_{Y|X})$, hence
for each $a\in \mathcal A_i$:
$$ p_e(a)\leq
 2^{n(E_{r}(R_i,W_{Y|X})-\epsilon_n)}.$$
\vspace{0.1 in}
 This channel coding theorem as Csisz\'{a}r put it, the ``main result of this paper''
 in~\cite{Csiszar:80}. We  use this theorem directly in the proof of
 the lower bound in Proposition~\ref{prop:both} and further modify it to show
the lower bound in Theorem~\ref{theorem:main}.

%
%
%


 \section{Joint source-channel coding error exponent with both decoder
 and encoder side-information}\label{sec.both}
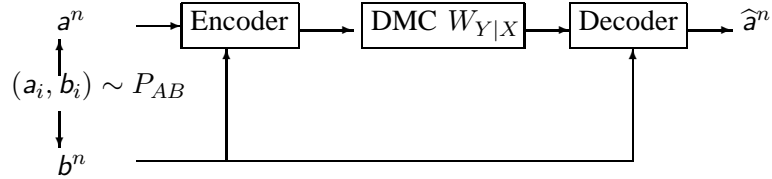
\begin{figure}[htbp]
\setlength{\unitlength}{0.6mm}
\begin{picture}(100,45)(-60,0)

\put(40,30){\line(1,0){26}} \put(40,40){\line(1,0){26}}
\put(40,30){\line(0,1){10}} \put(66, 30){\line(0,1){10}}
\put(42,34){Encoder }

\put(80,30){\line(1,0){36}} \put(80,40){\line(1,0){36}}
\put(80,30){\line(0,1){10}} \put(116, 30){\line(0,1){10}}
\put(82,34){DMC $W_{Y|X}$}

\put(126,30){\line(1,0){26}} \put(126,40){\line(1,0){26}}
\put(126,30){\line(0,1){10}} \put(152, 30){\line(0,1){10}}
\put(128,34){Decoder }

\put(116, 34){\vector(1,0){10}} \put(66, 34){\vector(1,0){12}}
\put(152, 34){\vector(1,0){10}}

 \put(30, 5){\line(1,0){110}}
\put(140, 5){\vector(0,1){25}} \put(50, 5){\vector(0,1){25}}

\put(162,34){ $\widehat{\rva}^n$}

\put(30, 35){\vector(1,0){10}}

\put(10, 34){ ${\rva}^n$}\put(10, 2){ ${\rvb}^n$}

\put(0, 20){ $(\rva_i,\rvb_i)\sim P_{AB}$}

\put(13,24){\vector(0,1){8}} \put(13,16){\vector(0,-1){8}}
 \end{picture}
     \caption[]{Source coding with both decoder  \textbf{and encoder}  side-information}
     \label{fig.both_side}
 \end{figure}

As a warmup to the more interesting scenario where the
side-information is not known to the encoder, we present the
upper/lower bounds when both the encoder and the decoder know the
side-information. This setup is shown in Figure~\ref{fig.both_side}.

The error probability of the coding system is, similar
to~(\ref{eqn.errorP_def}):
\begin{eqnarray}
\Pr(\rva^n \neq \widehat \rva^n(\rvb^n, \rvy^n))=\sum_{a^n, b^n}
P_{AB}(a^n, b^n)\sum_{y^n} W_{Y|X}(y^n| x^n(a^n, b^n)) 1(a^n \neq
\widehat a^n (b^n, y^n))\label{eqn.errorPboth_def}.
\end{eqnarray}
The error exponent of this setup is denoted by $E_{both}(P_{AB},
W_{Y|X})$  which is defined in the same way as $E_{}(P_{AB},
W_{Y|X})$ in~(\ref{eqn.EE_def}). The difference is that the encoder
observes both source $\sva^n$ and the side-information $\svb^n$,
hence the output of the encoder is a function of both: $x^n(\sva^n,
\svb^n)$. So obviously, $E_{both}(P_{AB}, W_{Y|X})$ is not smaller
than $E_{}(P_{AB}, W_{Y|X})$.

Comparing~(\ref{eqn.errorPboth_def})
and~(\ref{eqn.errorPAonly_def}), we can see the connections between
joint source-channel coding with both decoder and encoder side
information and joint source-channel coding. Knowing the side
information $b^n$, the joint source channel coding with both encoder
and decoder side information problem is essentially a channel coding
problem with messages distributed on $\mathcal A^n$ with a
distribution $P_{A|B}(\rva^n|b^n)$. Hence we can extend the
 results for joint source-channel coding error
exponent~\cite{Csiszar:80}. We summarize the bounds on
$E_{both}(P_{AB}, W_{Y|X})$ in the following
proposition.\vspace{0.1in}

\begin{proposition}{ Lower and upper bound on $E_{both}(P_{AB}, W_{Y|X})$} \label{prop:both}
\begin{eqnarray}
&&E_{both}(P_{AB}, W_{Y|X})\leq \min_{R}
 \{e_U(R, P_{AB}) +E_{sp}(R,W_{Y|X})\}\nonumber\\
&&E_{both}(P_{AB}, W_{Y|Z})\geq \min_{R}  \{e_U(R, P_{AB})
+E_{r}(R,W_{Y|X})\}
\end{eqnarray}
Not explicitly stated, but it should be clear that the range of $R$
is $(0,\log_2|\cal A|)$.
\end{proposition} \vspace{0.1in}

 \proof see Appendix~\ref{sec.appendix.both}. Because $E_{both}(P_{AB},
W_{Y|X})$ is no smaller than $E_{}(P_{AB}, W_{Y|X})$, so the lower
bound of $E_{}(P_{AB}, W_{Y|X})$ in Theorem~\ref{theorem:main} is
also a lower bound for $E_{both}(P_{AB}, W_{Y|X})$. However, in the
appendix, we give a simple proof of the lower bound on
$E_{both}(P_{AB}, W_{Y|X})$ which is a   corollary of Theorem~5
in~\cite{Csiszar:80}.  \hfill $\square$

 \vspace{0.1in}

Comparing the lower and the upper bounds for the case with both
encoder and decoder side-information, we can easily see that if
$R^*$ minimizes $ \{e_U(R, P_{AB}) +E_{r}(R,W_{Y|X})\}$ and
$E_{sp}(R^*,W_{Y|X})=E_{r}(R^*,W_{Y|X})$, then the upper bound and
the lower bound match. Hence,
\begin{eqnarray}
 E_{both}(P_{AB}, W_{Y|X})=e_U(R^*, P_{AB})+E_{r}(R^*,W_{Y|X})\label{eqn.both_complete}.
\end{eqnarray}

In this case $E_{both}(P_{AB}, W_{Y|X})$ is completely
characterized.

 \section{ joint source-channel  error exponents with only decoder side
 information}\label{sec.mainresult}
We study the more interesting problem where only decoder knows the
side-information in this section. We first give a lower and an upper
bound on the error exponent of joint source-channel coding with
decoder only side-information. The result is summarized in the
following Theorem.

 \vspace{0.1in}
\begin{theorem}{Lower and upper bound on the joint source channel coding with decoder side-information
only, as setup in Figure~\ref{fig.setup},
 error exponent:}\label{theorem:main} For the error probability $\Pr(\rva^n \neq \widehat \rva^n(\rvb^n, \rvy^n))$ and error
 exponent $E(P_{AB}, W_{Y|X})$
 defined in~(\ref{eqn.errorP_def}) and~(\ref{eqn.EE_def}), we have the following
 lower and upper
 bounds:
\begin{eqnarray}
&& E(P_{AB}, W_{Y|X})\geq\label{eqn.lowerbound}
\\
&&\min_{Q_A}\max_{S_X(Q_A)}\min_{Q_{B|A}, V_{Y|X}}
\{D(Q_{AB}\|P_{AB})+ D(V_{Y|X}\| W_{Y|X}|S_X(Q_A))
 +|I(S_X(Q_{A});
V_{Y|X})- H(Q_{A|B} )|^+\} \nonumber\\
&& E(P_{AB}, W_{Y|X})\leq\label{eqn.upperbound}
\\
&&\min\limits_{Q_A}\max\limits_{S_X(Q_A)}\min\limits_{Q_{B|A},V_{Y|X}:
I(S_X(Q_A);V_{Y|X})< H(Q_{A|B})}\{D(Q_{AB}\|P_{AB})+D(V_{Y|X}\|
W_{Y|X}|S_X(Q_A)) \}\nonumber
\end{eqnarray}\end{theorem}
\vspace{0.1in}

\proof The main technical tool used here is the method of types. For
the lower bound we propose a joint coding scheme for the joint
source channel coding with side information problem. This scheme is
a modification of the coding scheme first proposed
in~\cite{Csiszar:80}. However, we cannot directly use the channel
coding Theorem 5 in~\cite{Csiszar:80} because of the presence of the
side information. In essence, we have to study a more complicated
case using the method of types.  Details see
Appendix~\ref{appendix.proof_main}. \hfill $\square$

To simplify the expressions of the lower and upper bounds and later
give a sufficient condition for these two bounds to match, we
introduce the ``digital interface'' $R$ and have the following
corollary.

\vspace{0.1in}
\begin{corollary}{upper and lower bounds on  $E_{ }(P_{AB}, W_{Y|X})$ with ``digital interface'' $R$}\label{cor.digital}
 \begin{eqnarray}
&&E_{ }(P_{AB}, W_{Y|X})\leq \min_{Q_A}\max_{S_X(Q_A)}\min_{R}\{
e_U(R, P_{AB}, Q_A)+ E_{sp}(R,
S_X(Q_A), W_{Y|X})\} \label{eqn.digitalupper}\\
&&E_{ }(P_{AB}, W_{Y|Z})\geq \min_{Q_A}\max_{S_X(Q_A)}\min_{R}\{
e_U(R, P_{AB}, Q_A)+ E_r(R, S_X(Q_A), W_{Y|X})\}
\label{eqn.digitallower}
\end{eqnarray}
where $E_r(R, S_X(Q_A), W_{Y|X})$ is the standard random coding
error exponent for channel $W_{Y|X}$ at rate $R$ with input
distribution $S_X(Q_A)$ defined in~(\ref{eqn.channel_random_ee}),
while $e_U(R, P_{AB}, Q_A)$ is a peculiar source coding with
side-information error exponent for source $P_{AB}$ at rate $R$,
where the empirical source distribution is fixed at $Q_A$. That is
for $Q_A$
\begin{eqnarray}
e_U(R, P_{AB}, Q_A)
 &\triangleq& \min_{Q_{B|A}:  H(Q_{A|B} )\geq  R}
D(Q_{AB}\|P_{AB})\label{eqn.define_eu}
\end{eqnarray}

\end{corollary}

\vspace{0.1in} \proof The proof is in
Appendix~\ref{sec.appendix.digial}. \hfill $\square$

\vspace{0.1in}

With the simplified expression of the lower and upper bounds in
Corollary~\ref{cor.digital}, we can  give a game theoretic
interpretation of the bounds. And more importantly, we present some
sufficient conditions for the two bounds to match.

\subsection{A game theoretic interpretation of the bounds}

The lower and upper bounds established in
Corollary~\ref{cor.digital} clearly have a game theoretic
interpretation. This is a two player zero sum game. The first player
is ``nature'', the second player is the coding system, the payoff
from ``nature'' to the coding system is the bounds on the error
exponents in Corollary~\ref{cor.digital}. ``Nature'' chooses the
 marginal of the source $Q_A$ (observable to the coding
system) and  $R$ which is essentially the side information $Q_{B|A}$
and the channel behavior $V_{Y|X}$ (non-observable to the coding
system). The coding system choose $S_X(Q_A)$ after observing $Q_A$.
Hence in this game, the ''nature'' has two moves, the first move on
$Q_A$ and the last move on  $R$ which is essentially  $Q_{B|A}$ and
$V_{Y|X}$, while the coding system has the middle move on
$S_X(Q_A)$.

Comparing Corollary~\ref{cor.digital} for joint source-channel
coding with decoder side information and the classical joint
source-channel coding error exponent~\cite{Csiszar:80}
in~(\ref{eqn.jsc_csiszar}), it is desirable to have a sufficient
condition that the lower bound and the upper bound match, i.e. the
complete characterization as that in~(\ref{eqn.jsc_complete}).  It
is simpler for the case in~(\ref{eqn.jsc_csiszar}) since all is
needed is that the sphere backing bound and the random coding bound
to match at the critical rate $R^*$ as discussed in
Section~\ref{sec.jsc}. However, for the two bounds in
Corollary~\ref{cor.digital}, it is not clear what the conditions are
such that these two bounds match.   Suppose that the solution of the
game~(\ref{eqn.digitalupper}) is $(Q^u_A, S^u_X(Q_A), R^u )$ and
solution of the game~(\ref{eqn.digitallower}) is $(Q^l_A,
S^l_X(Q_A), R^l )$. An obvious sufficient condition for the two
bounds match is as follows:
\begin{eqnarray}
(Q^l_A, S^l_X(Q_A), R^l ) = (Q^u_A, S^u_X(Q_A), R^u ) \mbox{ and }
E_r(R^u, S^u_X(Q_A), W_{Y|X})=E_{sp}(R^u, S^u_X(Q_A),
W_{Y|X})\label{eqn.bad_suffi}
\end{eqnarray}
This  condition is hard to verify for \textit{any} source channel
pairs. In the next section, we try to simplify the condition under
which these two bounds match for a class of channels.

\subsection{A sufficient condition to reduce  $\min\{\max\{\min\{\cdot\}\}\}$ to
 $\min\{\cdot\}$}

The difficulty in studying the bounds in Corollary~\ref{cor.digital}
is that the $\min$ and $\max$ operators are nested.  The problem
will be simplified if we can change the order of the $\min$ and
$\max$ operators.

\begin{corollary}\label{cor.sufficient} For symmetric channels $W_{Y|X}$ defined
on Page 94 in~\cite{Gallager}, this includes the binary symmetric
and binary erasure channels, where the input distribution $S_X$ to
maximize the random coding error exponent $E_r(R, S_X, W_{Y|X})$ is
uniform on $\mathcal X$, or for more general channels\footnote{For
example, a channel consisted of parallel symmetric channels.}, where
the input distribution $S_X$ to maximize the random coding error
exponent $E_r(R, S_X, W_{Y|X})$ is the same for all $R$, then the
upper and lower bounds in Theorem~\ref{theorem:main} and
Corollary~\ref{cor.digital} can be further simplified to the
following forms:
 \begin{eqnarray}
&&E_{ }(P_{AB}, W_{Y|X})\leq \min_{R}
 \{e_U(R, P_{AB}) +E_{sp}(R,W_{Y|X})\}\label{eqn.mainupper}\\
&&E_{ }(P_{AB}, W_{Y|Z})\geq \min_{R}  \{e_U(R, P_{AB})
+E_{r}(R,W_{Y|X})\}\label{eqn.mainlower}
\end{eqnarray}
{\em Note: in this case, the upper and lower bounds for $E_{
}(P_{AB}, W_{Y|X})$ is the same as those for $E_{both }(P_{AB},
W_{Y|X})$ in Proposition~\ref{prop:both}. More discussions see
Section~\ref{sec.discussion}.\em}

\end{corollary}
\vspace{0.1in}

\proof An important property for symmetric channels is that the
input distribution that maximizes the random coding error exponent
is constant for all rate $R$, hence the inner $\max\min\{\cdot\}$ is
equal to $\min\max\{\cdot\}$, i.e.

\begin{eqnarray}
E(P_{AB}, W_{Y|X}) &\geq& \min_{Q_A}\max_{S_X(Q_A)}\min_{R}\{ e_U(R,
P_{AB}, Q_A)+ E_{r}(R, S_X(Q_A), W_{Y|X})\} \nonumber\\
&=& \min_{Q_A}\min_{R}\max_{S_X(Q_A)}\{ e_U(R,
P_{AB}, Q_A)+ E_{r}(R, S_X(Q_A), W_{Y|X})\} \nonumber\\
&=& \min_{Q_A} \min_{R}\{ e_U(R,
P_{AB}, Q_A)+ E_{r}(R,  W_{Y|X})\} \label{eqn.obvious0}\\
&=& \min_{R} \{\min_{Q_A}\{ e_U(R,
P_{AB}, Q_A)\}+ E_{r}(R,  W_{Y|X})\} \nonumber\\
&=& \min_{R} \{ e_U(R, P_{AB})+ E_r(R,
W_{Y|X})\}\label{eqn.obvious1}
\end{eqnarray}
where~(\ref{eqn.obvious0}) follows the definition of random coding
bound in~(\ref{eqn.erandom}) and~(\ref{eqn.obvious1}) follows the
obvious equality:
\begin{eqnarray}
\min_{Q_A} e_U(R, P_{AB}, Q_A)= \min_{Q_{AB}:  H(Q_{A|B} )\geq R}
D(Q_{AB}\|P_{AB})= e_U(R, P_{AB}).\nonumber
\end{eqnarray}

The upper bound in~\ref{eqn.mainupper} is trivial by noticing that
$\max\min\{\cdot\}\leq \min\max\{\cdot\}$~\cite{Convex}, hence:
\begin{eqnarray}
E(P_{AB}, W_{Y|X}) &\leq& \min_{Q_A}\max_{S_X(Q_A)}\min_{R}\{ e_U(R,
P_{AB}, Q_A)+ E_{sp}(R, S_X(Q_A), W_{Y|X})\} \nonumber\\
&\leq& \min_{Q_A}\min_{R}\max_{S_X(Q_A)}\{ e_U(R,
P_{AB}, Q_A)+ E_{sp}(R, S_X(Q_A), W_{Y|X})\} \nonumber\\
&=& \min_{Q_A} \min_{R}\{ e_U(R,
P_{AB}, Q_A)+ E_{sp}(R,  W_{Y|X})\} \nonumber\\
&=& \min_{R} \{\min_{Q_A}\{ e_U(R,
P_{AB}, Q_A)\}+ E_{sp}(R,  W_{Y|X})\} \nonumber\\
&=& \min_{R} \{ e_U(R, P_{AB})+ E_{sp}(R,  W_{Y|X})\}
\label{eqn.obvious2}
\end{eqnarray}
Corollary~\ref{cor.sufficient} is proved. \hfill$\square$
\vspace{0.1in}

 With this corollary proved, we can give a
 sufficient condition under which the lower bound and upper bound
 match similar to that for the joint source-channel coding case in
 Section~\ref{sec.jsc}. More discussions see Section~\ref{sec.discussion}.

\subsection{Why it is hard to generalize Corollary~\ref{cor.sufficient} to non-symmetric channels?}
Whether $\max\limits_{S_X(Q_A)}\min\limits_{R}\{ e_U(R, P_{AB},
Q_A)+ E_{r}(R, S_X(Q_A), W_{Y|X})\}$ is equal to
$\min\limits_{R}\max\limits_{S_X(Q_A)}\{ e_U(R, P_{AB}, Q_A)+
E_{r}(R, S_X(Q_A), W_{Y|X})\}$  is not obvious for general
(non-symmetric) channels. A sufficient condition of the existence of
a unique saddle point hence the equality is known as the Sion's
Theorem~\cite{Sion} which states that:
\begin{eqnarray}
\max_{\mu\in \mathcal M}\min_{\nu\in \mathcal N} f(\mu,
\nu)=\min_{\nu\in \mathcal N} \max_{\mu\in \mathcal M}f(\mu, \nu)
\end{eqnarray}
if $\mathcal M $ and $\mathcal N$ are convex, compact spaces and $f$
a quasi-concave-convex (definitions see~\cite{Convex}) and
continuous function on $\cal M \times N$. For the function of
interest,:
\begin{eqnarray}
&&\max\limits_{S_X(Q_A)}\min\limits_{R}\{ e_U(R, P_{AB}, Q_A)+
E_{r}(R, S_X(Q_A), W_{Y|X})\}\label{eqn.hard}.
\end{eqnarray}
We examine the sufficient condition under which a unique equilibrium
exists, according to the Sion's Theorem.  First, $e_U(R, P_{AB},
Q_A)+ E_{sp}(R, S_X(Q_A), W_{Y|X})$ is quasi-convex in $R$ because
both $e_U(R, P_{AB}, Q_A)$ and $E_{sp}(R, S_X(Q_A), W_{Y|X})$ are
convex, hence quasi-convex in $R$. However, (\ref{eqn.hard}) is not
quasi concave on $S_X(Q_A)$:
$$E_{r}(R, S_X(Q_A), W_{Y|X}) =\inf_{V_{Y|X}}D(V_{Y|X}\|W_{Y|X} | S_X(Q_A))+
|I(V_{Y|X};S_X(Q_A))-R|^+,$$
notice that the first term is linear in $S_X(Q_A)$, the second term
is quasi-concave but not concave. But the sum of a linear function
and a quasi-concave function might not be quasi-concave.  This shows
that the $\min\max$  theorem cannot be established by using the
Sion's Theorem. This does not mean that the $\min\max$ theorem
cannot be proved. However for a non quasi-concave function that may
have multiple peaks,   $\min\max\{\cdot\}$  is  not necessarily
equal to
 $\max\min\{\cdot\}$.

\begin{figure}[ht]
\begin{minipage}[b]{0.45\linewidth}
\centering
\includegraphics[scale=0.95]{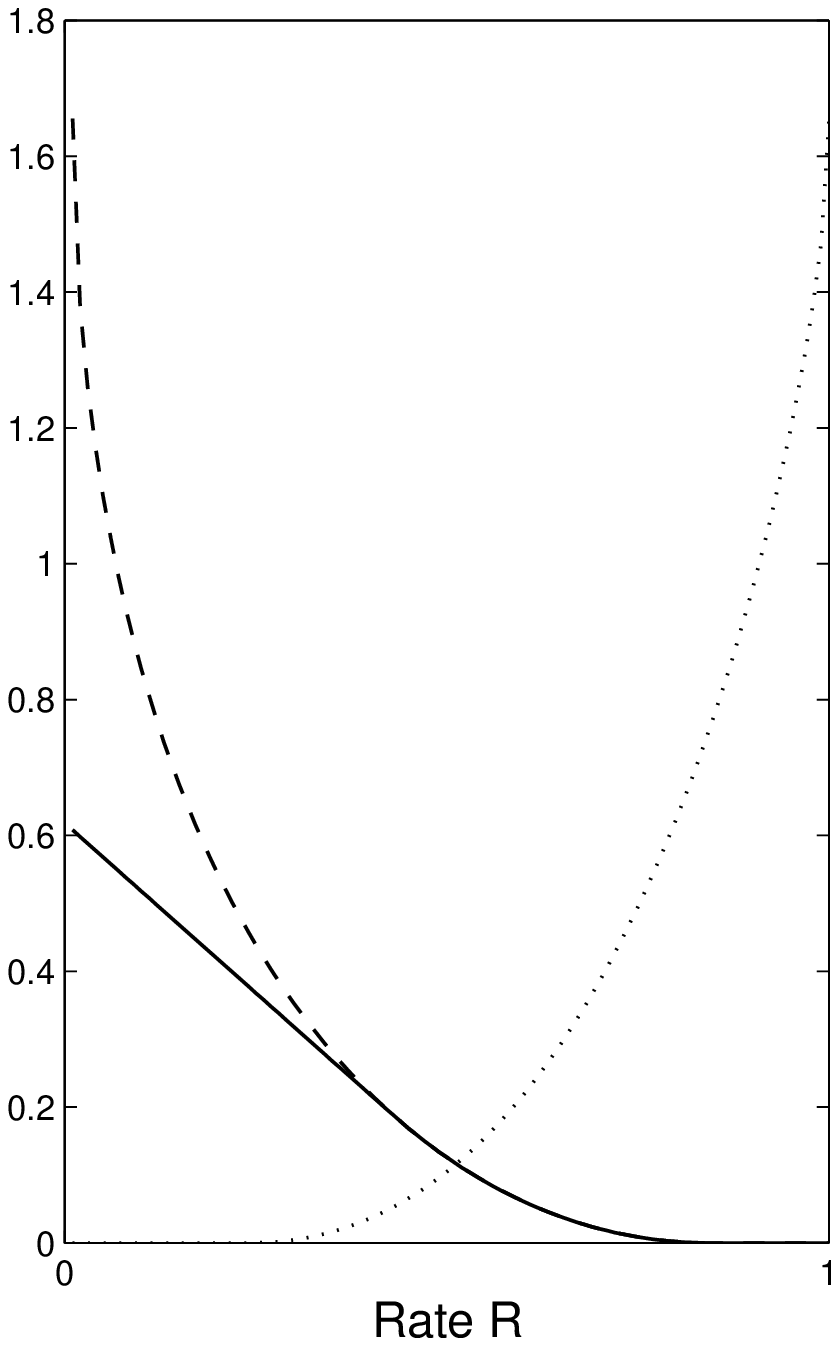}
\caption{ The upper bound on source coding with side-information
error exponent $ e_{U}(R, P_{AB})$ is the dotted line. The random
coding bound $E_{r}(R,W_{Y|X})$ and sphere packing bound
$E_{sp}(R,W_{Y|X})$ for channel coding error exponents are the solid
line and the dashed line respectively. } \label{fig:figure1}
\end{minipage}
\hspace{0.5cm}
\begin{minipage}[b]{0.45\linewidth}
\centering
\includegraphics[scale=0.95]{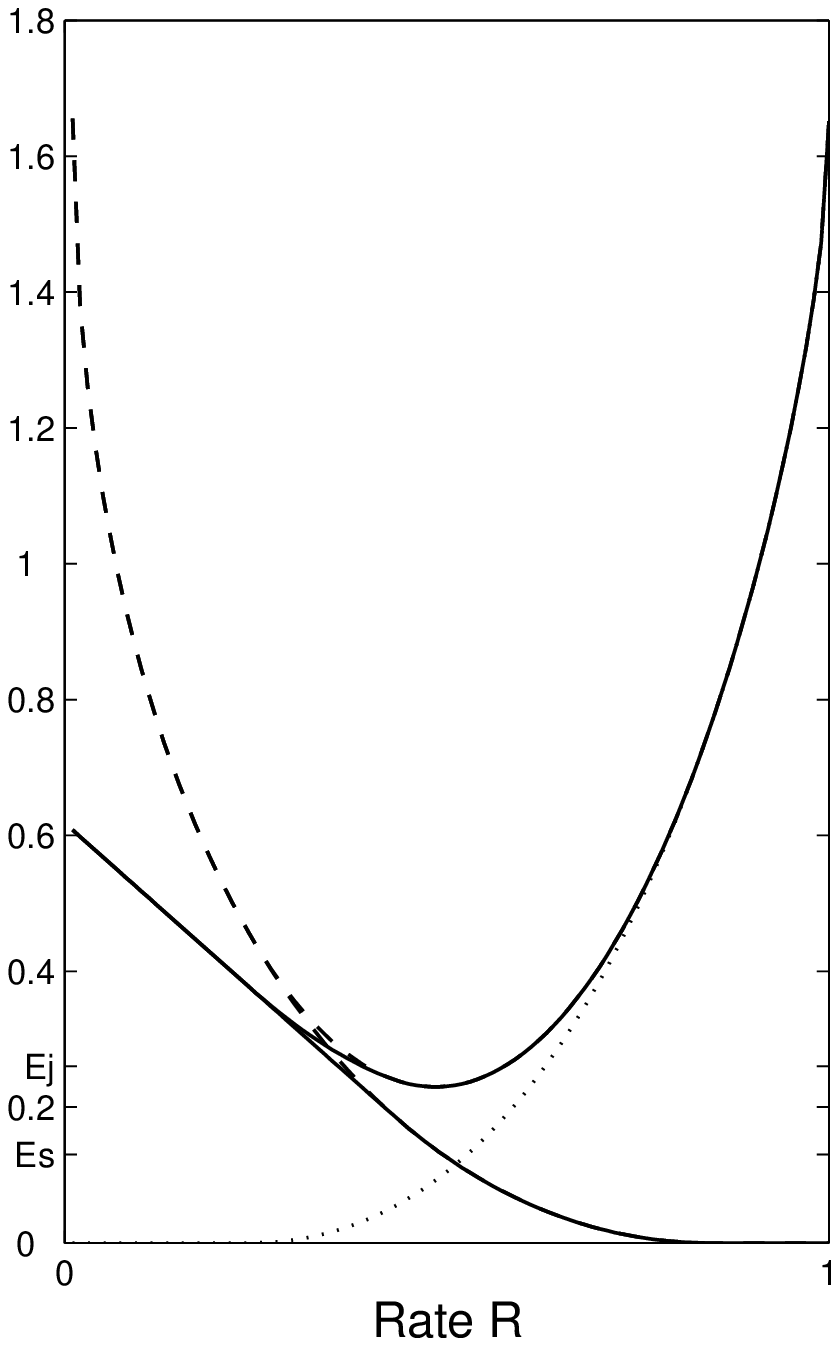}
\caption{$e_U(R, P_{AB}) +E_{sp}(R,W_{Y|X})$ and $e_U(R, P_{AB})
+E_{r}(R,W_{Y|X})$ are added to Figure~\ref{fig:figure1} in dashed
line and solid line respectively, they match at the minimal point
hence the joint source-channel coding with decoder side-information
error exponent is completely determined as $E_{ }(P_{AB},
W_{Y|X})=E_j$ And $E_s$ is the separate coding error exponent
$E_{separate}(P_{AB}, W_{Y|X})$ defined in~(\ref{eqn.sep}).}
\label{fig:figure2}
\end{minipage}

\end{figure}

\section{``Almost'' complete characterization of $E_{ }(P_{AB},
W_{Y|X})$ for symmetric channels}\label{sec.discussion}

The sufficient condition in Corollary~\ref{cor.sufficient} is
important, since binary symmetric and binary erasure channels are
among the most well studied discrete memoryless channels. We further
discuss the implications of the ``almost'' complete characterization
of $E_{ }(P_{AB}, W_{Y|X})$ for symmetric channels.

First we give an example shown in Figure~\ref{fig:figure1} and
Figure~\ref{fig:figure2}. The source $\rva$ is a Bernoulli $0.5$
random variable  and the joint distribution has the distribution
\begin{eqnarray}
P_{AB}=\left\{ \begin{array}{cc}
  0.50 & 0.00 \\
  0.05& 0.45 \\
  \end{array} \right\}
\end{eqnarray}
The channel $W_{Y|X}$ is a binary symmetric channel with cross rate
$0.025$. The channel coding error exponent  bounds
$E_{r}(R,W_{Y|X})$ and $E_{sp}(R,W_{Y|X})$ and the source coding
with decoder side-information upper bound $ e_{U}(R, P_{AB})$ are
plotted in Figure~\ref{fig:figure1}. The channel coding bound match
while $R\geq R_{cr} $, where $R_{cr}$ is defined
in~\cite{Gallager}.\\ {\em Note: the lower bound of the source
coding with side information error exponent $ e_{L}(R, P_{AB})$ is
not plotted in the figure.\em}

In Figure~\ref{fig:figure2}, we add both the lower and upper bounds
on the joint source channel coding with decoder side information to
the plot in Figure~\ref{fig:figure1}. For this source channel pair
$P_{AB}$ and $W_{Y|X}$, we have a complete characterization of
$E_{both}(P_{AB}, W_{Y|X})$ because the channel is symmetric and the
two bounds match at the minimal point, i.e. the two curves: $e_U(R,
P_{AB}) +E_{sp}(R,W_{Y|X})$ and $e_U(R, P_{AB}) +E_{r}(R,W_{Y|X})$
match at the minimal point as shown in Figure~\ref{fig:figure2}. The
value of the minimum is $E_j$ shown in Figure~\ref{fig:figure2}.

\subsection{Encoder side information often does not help}

 Similar to Proposition~\ref{prop:both}, we can see the
conditions under which we can give a complete characterization of
the joint source channel coding with decoder only side information
error exponent $E(P_{AB}, W_{Y|X})$. If $R^*$ minimizes $ \{e_U(R,
P_{AB}) +E_{r}(R,W_{Y|X})\}$ and
$E_{sp}(R^*,W_{Y|X})=E_{r}(R^*,W_{Y|X})$, then the upper bound and
the lower bound match. Hence:
\begin{eqnarray}
 E(P_{AB}, W_{Y|X})=e_U(R^*, P_{AB})+E_{r}(R^*,W_{Y|X})\label{eqn.si_complete}.
\end{eqnarray}

Comparing Corollary~\ref{cor.sufficient} and
Proposition~\ref{prop:both}, we bound the error exponent with or
without decoding side-information by the same lower and upper
bounds. This does not mean that $E_{ }(P_{AB},
W_{Y|Z})=E_{both}(P_{AB}, W_{Y|Z})$ always holds. But if the lower
bound and upper bound match, which is shown in
Figure~\ref{fig:figure2}, then we have:
\begin{eqnarray}
E_{ }(P_{AB}, W_{Y|Z})=E_{both}(P_{AB}, W_{Y|Z})=e_U(R^*,
P_{AB})+E_{r}(R^*,W_{Y|X}).
\end{eqnarray}
where $R^*$ minimizes $e_U(R, P_{AB})+E_{r}(R,W_{Y|X})$ and $R^*>
R_{cr}$. This is another example for block coding where knowing
side-information does not help increase the error exponent. In the
contrary, as discussed in~\cite{OurSIPaper}, in the delay
constrained setup, there is a penalty for not knowing the
side-information even if the channel is noiseless.

\subsection{Separate coding is strictly sub-optimal}
An obvious coding scheme for the problem in Figure~\ref{fig.setup}
is to implement a separate coding scheme. A source encoder first
encodes the source sequence $\sva^n$ into a rate $R$, where $R$ is
determined later,  bit stream $c^{nR}(\sva^n)$ then an independent
channel encoder encodes the bits $c^{nR}$ into channel inputs
$\svx^n$. The channel decoder first decodes the channel output
$\svy^n$ into bits $\widehat c^{nR}$ and then the independent source
decoder reconstructs $\widehat a^n$ from $\widehat c^{nR}$ and side
information $b^n$. This is a separate coding scheme with outer
source  with side information coding and inner channel coding, both
at rate $R$. If both coding are random coding that achieves the
random coding error exponents for both source coding and channel
coding respectively. The union bound of the error probability is as
follows:
\begin{eqnarray}
\Pr(\rva^n \neq \widehat \rva^n(\rvb^n, \rvy^n))&=& \Pr(\rvc^{nR}
\neq \widehat \rvc^{nR}(\rvy^n))+\Pr(\rva^n\neq \widehat
\rva(\widehat
\rvc^{nR}(\rvy^n),\rvb^n),\rvc^{nR}=\rvc^{nR}(\rvy^n))\label{eqn.unionbound1}\\
&\leq& \Pr(\rvc^{nR} \neq \widehat \rvc^{nR}(\rvy^n))+\Pr(\rva^n\neq
\widehat \rva(\widehat
\rvc^{nR}(\rvy^n),\rvb^n)|\rvc^{nR}=\rvc^{nR}(\rvy^n))\label{eqn.unionbound2}\\
&\leq & 2^{- n (E_{r}(R, W_{Y|X})- \epsilon_n^1)} + 2^{- n (e_{L}(R,
P_{AB})- \epsilon_n^2)}\label{eqn.unionbound3}
\end{eqnarray}
where $\epsilon_n^1$ and $\epsilon_n^2$ converges to zero as $n$
goes to infinity.  (\ref{eqn.unionbound1}) follows the union bound
argument that a decoding error occurs if either the inner channel
coding fails or the outer source coding fails.
(\ref{eqn.unionbound2}) is true because conditional probability is
large or equal to joint probability. Finally (\ref{eqn.unionbound3})
is true because both the outer source coding and inner channel
coding achieve the random coding error exponents.
From~(\ref{eqn.unionbound3}) and that we can optimize the digital
interface rate $R$ between the channel coder and source coder, we
know that the separate coding error exponent is
\begin{eqnarray}
\max_{R}\{\min \{E_{r}(R, W_{Y|X}),e_{L}(R, P_{AB}\}\}\triangleq
E_{separate}(P_{AB}, W_{Y|X})\label{eqn.sep}
\end{eqnarray}

This separate coding scheme is also discussed for joint source
channel coding in~\cite{Csiszar:80}. A similar bound is drawn. We
next show why the separate coding error exponent
$E_{separate}(P_{AB}, W_{Y|X})$ is in general strictly smaller than
the lower bound of $E(P_{AB}, W_{Y|X})$ in~(\ref{eqn.mainlower}).

First, obviously, $E_{separate}(P_{AB}, W_{Y|X})\leq
\max\limits_{R}\{\min \{E_{r}(R, W_{Y|X}),e_{U}(R, P_{AB})\}\}$.
Secondly  $\{E_{r}(R, W_{Y|X})$ is monotonically decreasing,
$e_{U}(R, P_{AB})$ is monotonically increasing, and both are
continuous and convex as shown in Figure~\ref{fig:figure2}. This
means that  for rate $\bar R$ such that  $E_{r}(\bar R,
W_{Y|X})=e_{U}(\bar R, P_{AB})$:
$$E_{separate}(P_{AB}, W_{Y|X})=
E_{r}(\bar R, W_{Y|X})=e_{U}(\bar R, P_{AB})$$
 Now let $R^*$ be the rate to minimize $\{e_U(R, P_{AB})
+E_{r}(R,W_{Y|X})\}$, i.e.$$E_{}(P_{AB}, W_{Y|X})\geq e_U(R^*,
P_{AB})+E_{r}(R^*,W_{Y|X}).$$

There are three scenarios. First if $R^*=\bar R$, then
$$E_{}(P_{AB}, W_{Y|X})\geq e_U(R^*, P_{AB})+E_{r}(R^*,W_{Y|X})= 2
E_{r}(\bar R,W_{Y|X})=2E_{separate}(P_{AB}, W_{Y|X}).$$

Secondly, if $R^*<\bar R$, $$E_{}(P_{AB}, W_{Y|X})\geq
E_{r}(R^*,W_{Y|X})>E_{r}(\bar R,W_{Y|X})=E_{separate}(P_{AB},
W_{Y|X}).$$

Finally if $R^*> \bar R$,
$$E_{}(P_{AB}, W_{Y|X})\geq
 e_U(R^*, P_{AB})>e_U(\bar R, P_{AB})=E_{separate}(P_{AB},
W_{Y|X}).$$

So in all cases, the joint source channel coding error exponent
$E_{}(P_{AB}, W_{Y|X})$ is strictly larger than the separate coding
error exponent $E_{separate}(P_{AB}, W_{Y|X})$. This is clearly
illustrated in Figure~\ref{fig:figure2}.\\

{\em  Note:   $E_{separate}(P_{AB}, W_{Y|X})$ is an achievable
separate coding error exponent from the obvious separate coding
scheme. What we prove is that this obvious one is strictly smaller
than the joint source-channel coding error exponent. This is similar
to the claim Csisz\'{a}r makes in~\cite{Csiszar:80}. It should be
clear that the upper bound of any separate source channel coding
error exponent is $\max_{R}\{\min \{E_{sp}(R, W_{Y|X}),e_{U}(R,
P_{AB}\}\} $ which is comparable to~(\ref{eqn.sep}). The proof
hinges on the complete transparency between the source coding and
channel coding, otherwise we have a joint coding schemes. A detailed
discussion is in~\cite{Zhong_phd}.\em}

\section*{Acknowledgement}
The   author  would like to thank Neri Merhav and Filippo
 Balestrieri for pointing out the Sion's theorem in~\cite{Sion}, and Fady Alajaji for
many helpful discussions on the technical results of this paper.

\section{Conclusions}

We study the joint source channel coding with decoder
side-information problem, with or without encoder side-information.
This is an extension of Csisz{\'a}r's joint source channel coding
error exponent problem in~\cite{Csiszar:80}. To derive the lower
bound, we use a novel joint source channel with decoder
side-information decoding scheme. We further investigate the
conditions under which the lower bounds and upper bounds match. A
game theoretic approach is applied to show the equivalence of the
lower and upper bound. This approach might be useful in simplifying
other error exponents with a cascade of min-max operators , for
example, the Wyner-Ziv coding error exponent recently studied
in~\cite{WZ_EE}.

\appendix
\subsection{Proof of upper and lower bounds on $E_{both}(P_{AB}, W_{Y|X})$}
\label{sec.appendix.both}

We prove Proposition~\ref{prop:both} in this section. The upper
bound and lower bounds are simple corollaries of the method of types
and Theorem~5 in~\cite{Csiszar:80} respectively. \vspace{0.1in}
\subsubsection{Upper bound}

 Consider a distribution $Q_{AB}$,  the joint
source channel encoder observes the realization of the source $(a^n,
b^n)$ with type $Q_{AB}$, for the case where the decoder knows the
side-information $b^n$. There are\footnote{Here $\epsilon^i_n$ goes
to zero as $n$ goes to infinity, $i=1,2,3$.} $2^{n
(H(Q_{A|B})-\epsilon^1_n)}$ many equally likely sequences $\in
\mathcal A^n$ conditional on $b^n$. These are the sequences with the
same joint probability with $b^n$ as the sequence $a^n$. Even
knowing the joint type $Q_{AB}$ (given by a genie) and the
side-information $b^n$, the decoder needs to guess the correct one
from the channel output $y^n$. This is a channel coding problem with
rate $H(Q_{A|B})-\epsilon^1_n$.

Now consider the channel input $x^n(a^n, b^n)$ where $b^n$ is the
side-information, notice that there are at most $(n+1)^{|\mathcal
X|}$ many different input types, there is a type $S_X(Q_{AB})$, such
that more than $(n+1)^{-|\mathcal X|}=2^{-n\epsilon^2_n}$ fraction
of the channel inputs given side-information $b^n$ and the joint
type of $(a^n, b^n)$ being $Q_{AB}$ have type $S_X(Q_{AB})$. For a
channel $V_{Y|X}$, such that the channel capacity of the channel
given the input distribution $S_X$ is smaller than $H(Q_{A|B})$,
i.e. $$I(S_X(Q_{AB}); V_{Y|X}) < H(Q_{A|B}),$$ then if the channel
$W_{Y|X}$ behaves like $V_{Y|X}$ with the code book with type
$S_X(Q_{AB})$, with high probably, the decoder cannot correctly
decide from one of the $2^{n H(Q_{A|B})}$ sequences. This is
guaranteed by the Blowing up Lemma~\cite{Csiszar} or see a detailed
proof in~\cite{Gallager_sphere}.

The probability that both the source behaves like $Q_{A|B}$ and the
channel behaves like $V_{Y|X}$ is
\begin{eqnarray}
2^{-n (D(Q_{AB}\|P_{AB})+D(V_{Y|X}\|
W_{Y|X}|S_X(Q_{AB}))-\epsilon^3_n)}.\label{eqn.bothupper1}
\end{eqnarray}

Notice that the source behavior $Q_{AB}$ and the channel behavior
$V_{Y|X}$ are arbitrary, as long as $ H(Q_{A|B})> I(S_X(Q_{AB});
V_{Y|X})$, we can upper bound the error exponent as follows:
\begin{eqnarray}
&& E_{both}(P_{AB}, W_{Y|Z})\nonumber\\
& \leq & \min_{Q_{AB}, V_{Y|Z}: H(Q_{A|B})> I(S_X(Q_{AB});
V_{Y|X})} \{D(Q_{AB}\|P_{AB})+D(V_{Y|X}\| W_{Y|X}|S_X(Q_{AB}))\}\label{eqn.bothupper2}\\
&=&\min_{R} \{\min_{Q_{AB}, V_{Y|Z}: H(Q_{A|B})>R
> I(S_X(Q_{AB});
V_{Y|X})} D(Q_{AB}\|P_{AB})+D(V_{Y|X}\| W_{Y|X}|S_X(Q_{AB}))\}\label{eqn.bothupper3}\\
&=&\min_{R} \{\min_{Q_{AB}: H(Q_{A|B})>R  } \{D(Q_{AB}\|P_{AB}) +
\min_{ V_{Y|Z}:  R > I(S_X(Q_{AB});
V_{Y|X})}  D(V_{Y|X}\| W_{Y|X}|S_X(Q_{AB}))\}\}\label{eqn.bothupper4}\\
&\leq &\min_{R} \{\min_{Q_{AB}: H(Q_{A|B})>R  } \{D(Q_{AB}\|P_{AB})
+ E_{sp}(R,W_{Y|X})\}\}\label{eqn.bothupper5}\\
&= &\min_{R } \{e_U(R, P_{AB})+E_{sp}(R,W_{Y|X})
\}\label{eqn.bothupper6}
%
\end{eqnarray}
(\ref{eqn.bothupper2}) is a direct consequence
of~(\ref{eqn.bothupper1}). In~(\ref{eqn.bothupper3}), we introduce
the ``digital interface'' $R$, the equivalence
in~(\ref{eqn.bothupper3}) and~(\ref{eqn.bothupper4})  should  be
obvious. (\ref{eqn.bothupper5}) and~(\ref{eqn.bothupper6})  are by
definitions of the channel coding and source coding error exponents.
\hfill $\square$ \vspace{0.1in}
\subsubsection{Lower bound}

  Given a side-information sequence $b^n$ which is known to
both the encoder and the decoder. We partition the source sequence
set $\mathcal A^n$ based on their joint type with $b^n$. The number
of joint types $m_n\leq (n+1)^{|\mathcal A||\mathcal B|}$ and denote
by $  Q_{AB}^i$, $i=1,2,... m_n$ the joint types. It should be clear
that the $Q_{AB}^i$'s here all have the same marginal distribution
as $b^n$.

$$\mbox{Let } \mathcal A_i(b^n)=\{a^n: (a^n, b^n)\in   Q_{AB}^i\},\ \  i =1,2,... m_n.$$
Obviously, $\mathcal A_i$'s form a partition of $\mathcal A^n$. And
each set has size $|A_i(b^n)|\leq 2^{nH(Q^i_{A|B})}$. Now we can
apply Theorem 5 of~\cite{Csiszar:80} as recited earlier: there
exists a channel code $f_0, \phi_0$, such that for each $a^n\in
\mathcal A_i(b^n)$, i.e. $(a^n, b^n)\in Q^i_{AB}$:
\begin{eqnarray}
p^{}_{e,b^n}(a^n)=\sum_{y^n} W_{Y|X}(y^n| x^n(a^n,b^n)) 1(a^n \neq
\widehat a^n (b^n, y^n))\leq
2^{-n(E_{r}(H(Q^i_{A|B}),W_{Y|X})-\epsilon_n)}\label{eqn.channel_both}.
\end{eqnarray}
The joint source channel coding error probability is hence:
\begin{eqnarray}
\Pr(\rva^n \neq \widehat \rva^n(\rvb^n, \rvy^n))&=&\sum_{a^n, b^n}
P_{AB}(a^n, b^n)\sum_{y^n} W_{Y|X}(y^n| x^n(a^n)) 1(a^n \neq
\widehat a^n (b^n, y^n))\nonumber\\
&=&\sum_{Q_{AB}}\sum_{(a^n, b^n)\in Q_{AB}} P_{AB}(a^n,
b^n)\sum_{y^n} W_{Y|X}(y^n| x^n(a^n,b^n)) 1(a^n \neq
\widehat a^n (b^n, y^n))\nonumber\\
&\leq &\sum_{Q_{AB}}\sum_{(a^n, b^n)\in Q_{AB}} P_{AB}(a^n,
b^n) 2^{-n(E_{r}(H(Q_{A|B}),W_{Y|X})-\epsilon_n)}\label{eqn.both1.1}\\
&\leq &\sum_{Q_{AB}}2^{-n D(Q_{AB}\|P_{AB})} 2^{-n(E_{r}(H(Q_{A|B}),W_{Y|X})-\epsilon_n)}\nonumber\\
&\leq & (n+1)^{|\mathcal A||\mathcal B|}\max_{Q_{AB}}\{2^{-n D(Q_{AB}\|P_{AB})} 2^{-n(E_{r}(H(Q_{A|B}),W_{Y|X})-\epsilon_n)}\}\nonumber\\
&\leq & 2^{-n(\min\limits_{Q_{AB}}\{
D(Q_{AB}\|P_{AB})+E_{r}(H(Q_{A|B}),W_{Y|X})\}-\epsilon'_n)}\label{eqn.both2.000}
\end{eqnarray}
(\ref{eqn.both1.1}) follows by substituting
in~(\ref{eqn.channel_both}) and the rest inequalities are by method
of types. $\epsilon'_n\rightarrow 0$, so we can lower bound the
error exponent as
\begin{eqnarray}
E_{both}(P_{AB}, W_{Y|Z}) & \geq & \min\limits_{Q_{AB}}\{
D(Q_{AB}\|P_{AB})+E_{r}(H(Q_{A|B}),W_{Y|X})\}\label{eqn.both2.00a}\\
&=&  \min_{R} \{\min\limits_{Q_{AB}:H(Q_{A|B})=R}\{
D(Q_{AB}\|P_{AB})+E_{r}(H(Q_{A|B}),W_{Y|X})\}\}\label{eqn.both2.00b}\\
&=&  \min_{R}\{ \min\limits_{Q_{AB}:H(Q_{A|B})=R}\{
D(Q_{AB}\|P_{AB})\}+E_{r}(R,W_{Y|X})\}\label{eqn.both2.1a}\\
&=&  \min_{R\geq H(P_{A|B})}\{ \min\limits_{Q_{AB}:H(Q_{A|B})=R}\{
D(Q_{AB}\|P_{AB})+E_{r}(R,W_{Y|X})\}\}\label{eqn.both2.1b}\\
&=&  \min_{R\geq H(P_{A|B})}\{ \min\limits_{Q_{AB}:H(Q_{A|B})\geq
R}\{
D(Q_{AB}\|P_{AB})+E_{r}(R,W_{Y|X})\}\}\label{eqn.both2.1c}\\
&=&  \min_{R\geq H(P_{A|B})}\{ e_U(R, P_{AB}) +E_{r}(R,W_{Y|X})
\}\label{eqn.both2.2a}\\
&=&  \min_{R}\{ e_U(R, P_{AB}) +E_{r}(R,W_{Y|X})
\}\label{eqn.both2.2b}
\end{eqnarray}
(\ref{eqn.both2.00a}) is a direct consequence
of~(\ref{eqn.both2.000}), in~(\ref{eqn.both2.00b}) we again
introduce the ``digital interface'' variable $R$.
(\ref{eqn.both2.1a}) and (\ref{eqn.both2.2a}) are by definitions of
$E_{r}(R,W_{Y|X})$ and $e_U(R, P_{AB})$ respectively.
(\ref{eqn.both2.1b}) is true because $E_{r}(R,W_{Y|X})$ is
monotonically increasing with $R$ and for $R<H(P_{B|A})$,
$$\min\limits_{Q_{AB}:H(Q_{A|B})=R}D(Q_{AB}\|P_{AB})\geq 0
=\min\limits_{Q_{AB}:H(Q_{A|B})=H(P_{A|B})}D(Q_{AB}\|P_{AB}).$$
(\ref{eqn.both2.1c}) is true because $D(Q_{AB}\|P_{AB})$ is convex
in $Q_{AB}$ and the global minimum is $Q^*_{AB}=P_{AB}$, but
$H(Q^*_{A|B})=H(P_{A|B})\geq R$ which means the minimum point is on
the boundary. Lastly~(\ref{eqn.both2.2b}) is because for $R<
H(P_{A|B})$, $e_U(R, P_{AB})$ is constant at $0$, while
$E_{r}(R,W_{Y|X})$ is monotonically increasing with $R$.
\hfill$\square$

\subsection{Lower and upper bounds on $E_{}(P_{AB},
W_{Y|X})$}\label{appendix.proof_main}

We give the proof of Theorem~\ref{theorem:main} here.

\subsubsection{Lower bound}  From the definition of the error exponent, we need to find a
encoding rule $x: \mathcal A^n \rightarrow \mathcal X^n$ and
decoding rule $\widehat a : \mathcal B^n \times \mathcal Y^n
\rightarrow \mathcal X^n$such chat the error probability :
\begin{eqnarray}
\Pr(\rva^n \neq \widehat \rva^n(\rvb^n, \rvy^n))=\sum_{a^n, b^n}
P_{AB}(a^n, b^n)\sum_{y^n} W_{Y|X}(y^n| x^n(a^n)) 1(a^n \neq
\widehat a^n (b^n, y^n))
\end{eqnarray}
is upper bounded by $2^{n(E-\epsilon_n)}$ where
$\epsilon_n\rightarrow 0$, where $E$ is the right hand side
of~(\ref{eqn.lowerbound}).


  We first describe the encoder and decoder, then prove that
this coding system achieves the lower bound.

The encoder only observes the source sequence $a^n$. For all those
sequences $a^n$ with type $Q_A$, the channel input is $x^n(a^n)$
that has type $S_X(Q_A)$, i.e. the channel input type only depends
on the type of the source, where $S_X(Q_A)$ is the distribution to
maximize the following exponent:
$$\min_{Q_{B|A}, V_{Y|X}}\{D(Q_{AB}\|P_{AB})+
D(V_{Y|X}\| W_{Y|X}|S_X(Q_A))+ |I(S_X(Q_{A}); V_{Y|X})- H(Q_{A|B}
)|^+\}  .$$

 The decoder observes both the side-information $b^n$ and the channel
output $y^n$, the decoder takes both the conditional entropy and
mutual information across the channel into account:
\begin{eqnarray}
\widehat a^n(b^n, y^n)=\argmax_{a^n}I(x^n(a^n); y^n)- H(a^n|b^n )
\end{eqnarray}

We next need to show that there exists such a encoder/decoder pair
that achieve the error exponent in~(\ref{eqn.lowerbound}). We also
use the method of random selection of codebooks. We denote by
$\mathcal C$ the set of the codebooks such that the codewords for
$a^n\in Q_A$ all have composition $S_X(Q_A)$. Obviously $\cal C$ is
finite, we let $ \zeta$ be the random variable uniformly distributed
on $\cal C$. We use codebook $c$ if $\zeta=c$, i.e. we use the
codebooks
  with equal probability. The most important property of this
  codebook distribution
  is the point-wise independence of the codewords, for all $a^n\in Q_A$ and
  $\tilde a^n\in \tilde Q_A$, for any two valid codewords $s^n \in S_X(Q_A)$ and
  $\tilde s^n\in S_X(\tilde Q_A)$ :
  \begin{eqnarray}
  \Pr^{\zeta}(x^n(a^n)= s^n, x^n(\tilde a^n)=\tilde s^n)=   \Pr^{\zeta}(x^n(a^n)= s^n)\Pr^{\zeta}( x^n(\tilde a^n)=\tilde s^n)
  =\frac{1}{|S_X(  Q_A)|}\frac{1}{|S_X(\tilde Q_A)|}
  \end{eqnarray}

We calculate the average error probability on the whole codebook set
$\cal C$. Write the average error probability as $p_e^n$, then first
we have:
\begin{eqnarray}
p_e^n=E(\Pr^\zeta(\rva^n \neq \widehat \rva^n(\rvb^n,
\rvy^n))=\frac{1}{|\mathcal C|}\sum_{c\in \mathcal C}\Pr^c(\rva^n
\neq \widehat \rva^n(\rvb^n, \rvy^n)),\label{eqn.def_lower}
\end{eqnarray}
where $E(\Pr\limits^\zeta(\rva^n \neq \widehat \rva^n(\rvb^n,
\rvy^n))$ is the expected error probability over all codebooks under
the codebook distribution $\zeta$.

For a fixed codebook $c\in \mathcal C$
\begin{eqnarray}
&&\Pr^c(\rva^n \neq \widehat \rva^n(\rvb^n, \rvy^n)) \nonumber\\
&=& \sum_{a^n, b^n} P_{AB}(a^n, b^n)\Pr^c(a^n \neq \widehat a^n(b^n,
\rvy^n) )\nonumber\\
&=&\sum_{Q_{AB}}\sum_{(a^n, b^n)\in Q_{AB}}\left(P_{AB}(a^n,
b^n)\Pr^c(a^n
\neq \widehat a^n(b^n, \rvy^n) )\right)\nonumber\\
&=&\sum_{Q_{AB}}\sum_{(a^n, b^n)\in Q_{AB}}\left(P_{AB}(a^n,
b^n)\sum_{y^n} W_{Y|X}(y^n|x^n(a^n))1^c(a^n
\neq \widehat a^n(b^n, y^n) )\right)\nonumber\\
&=&\sum_{Q_{AB}}\sum_{(a^n, b^n)\in
Q_{AB}}\nonumber\\
&&\left(P_{AB}(a^n, b^n)\sum_{V_{Y|X}}\hspace{0.1in}\sum_{y^n:
(x^n(a^n), y^n)\in S_X(Q_A)\times V_{Y|X}}
W_{Y|X}(y^n|x^n(a^n))1^c(a^n \neq \widehat a^n(b^n, y^n) )\right)
\label{eqn.lower_linearE1}
\end{eqnarray}
For  $(a^n, b^n)\in Q_{AB}$, so the source sequence $a^n$ has
marginal distribution $Q_A$, from the codebook generation we know
that the codeword $x^n(a^n)\in S_X(Q_A)$. For side-information
$b^n\in \mathcal B^n$, we partition $\mathcal A^n$ according to the
joint type with $b^n$:
$$Q_{\tilde AB}(b^n)=\{\tilde a^n\in \mathcal A^n: (\tilde  a^n, b^n) \in Q_{\tilde AB}\}.$$

 We  partition $S_X(Q_{\tilde A})$ according to the joint
distribution with $y^n$. For a joint distribution $U_{XY}$ s.t.
$U_X=S_X(Q_{\tilde A})$ and $y^n\in U_Y$:
$$U_{XY}(  Q_{\tilde A}, y^n)=\{ x^n \in S_X(Q_{\tilde A}): (x^n, y^n)\in   U_{XY}\}.$$

For   $(a^n,b^n)\in Q_{AB}$ and channel output $y^n\in \mathcal
Y^n$, s.t. $(x^n(a^n),y^n)\in V_{Y|X}$, a decoding error is made if
there exists a source sequence $\tilde  a^n \neq a^n$,s.t. $\tilde
a^n \in Q_{\tilde AB} (b^n)$ where $Q_{\tilde AB} $ may or may not
be $Q_{AB}$ and  the code word $ x^n(\tilde a^n)\in U_{XY}(y^n,
Q_{\tilde A})$, where $U_X=S_X(Q_{\tilde A})$ and $y^n\in U_Y$:
\begin{eqnarray}
&&I(x^n(\tilde a^n); y^n) - H(\tilde a^n|b^n ) \geq I(x^n(a^n);
y^n)-
H(a^n|b^n )\nonumber\\
i.e. && I_{U_{XY}}(X;Y) - H(Q_{\tilde A|B} ) \geq I(S_X(Q_{A});
V_{Y|X})- H(Q_{A|B} )\label{eqn.bounding_atypicality}
\end{eqnarray}

Now we can expand the indicator function
in~(\ref{eqn.lower_linearE1}) as follows, for a codebook $c$:

$1^c(a^n \neq \widehat a^n(b^n, y^n) )$
\begin{eqnarray}
&=&1^c\left(\exists \tilde a^n\neq a^n, s.t.  I(x^n(\tilde a^n),
y^n) - H(Q_{\tilde A|B} ) \geq
I(S_X(Q_{A}); V_{Y|X})- H(Q_{A|B} )\right)\nonumber\\
&\leq& \min\{1, \sum_{Q_{\tilde AB}, \ U_{XY}: I_{U_{XY}}(X,Y) -
H(Q_{\tilde A|B} ) \geq I(S_X(Q_{A}); V_{Y|X})- H(Q_{A|B}
)}1^c(\exists \tilde a^n\neq a^n \mbox{ and } \tilde a^n \in
Q_{\tilde AB}(b^n), \nonumber\\
&&   s.t. \ x^n(\tilde a^n)\in U_{XY}( Q_{\tilde A},
y^n)\}\label{eqn.indicator}
 \end{eqnarray}

Under the uniform codebook distribution $\zeta$, for $\tilde a^n
\neq a^n$,  $x^n( \tilde a^n)$ is uniformly distributed in
$S_X(Q_{\tilde A})$ independent of $x^n(a^n)$, so for all $Q_{\tilde
AB}$ and $U_{XY}$ with the proper marginals($b^n\in Q_B,$
 $U_X=S_X(Q_{\tilde A})$ and $y^n \in U_Y$) and satisfying~(\ref{eqn.bounding_atypicality}): \vspace{0.05in}

\begin{eqnarray}
&&E( 1(\exists \tilde a^n\neq a^n \mbox{ and } \tilde a^n \in
Q_{\tilde AB}(b^n),  s.t. \ x^n(\tilde a^n)\in U_{XY}( Q_{\tilde A},
y^n)))\nonumber\\
&=&\frac{1}{|\mathcal C|}\sum_{c\in \mathcal C} 1^c(\exists \tilde
a^n\neq a^n \mbox{ and } \tilde a^n \in Q_{\tilde AB}(b^n),  s.t. \
x^n(\tilde a^n)\in U_{XY}( Q_{\tilde A}, y^n))\nonumber\\
&=&\Pr^\zeta (\exists \tilde a^n\neq a^n \mbox{ and } \tilde a^n \in
Q_{\tilde AB}(b^n),  s.t. \ x^n(\tilde a^n)\in U_{XY}( Q_{\tilde A},
y^n))\nonumber\\
&\leq&|Q_{\tilde AB}(b^n)| \Pr^\zeta (  \ x^n(\tilde a^n)\in U_{XY}(
Q_{\tilde A}, y^n)|  \tilde a^n\neq a^n \mbox{
and } \tilde a^n \in Q_{\tilde AB}(b^n))\label{eqn.unionbound}\\
&=& |Q_{\tilde AB}(b^n)| \frac{|U_{XY}( Q_{\tilde A},
y^n)|}{|S_X(Q_{\tilde A})|}\label{eqn.uniform_distr}\\
&\leq& 2^{n\epsilon_n} 2^{nH(Q_{\tilde A|B})}\frac{2^{n H(U_{X|Y})
}}{2^{n H(U_{X})}}\label{eqn.method_of_type}\\
&=& 2^{-n(I_{U_{XY}}(X,Y) - H(Q_{\tilde A|B}
)-\epsilon_n)}\nonumber\\
&\leq & 2^{-n (I(S_X(Q_{A}); V_{Y|X})- H(Q_{A|B} )-\epsilon_n)}
\label{eqn.final_touch}
\end{eqnarray}

where $\epsilon_n\rightarrow 0$. (\ref{eqn.unionbound}) is by a
union bound argument. (\ref{eqn.uniform_distr}) is true because the
codeword $x^n(\tilde a^n) $ is  uniformly distributed in
$S_X(Q_{\tilde A})$. (\ref{eqn.method_of_type}) is by the method of
types. (\ref{eqn.final_touch}) is  true because the condition
in~(\ref{eqn.bounding_atypicality}) is satisfied.

Combining~(\ref{eqn.indicator}) and~(\ref{eqn.final_touch}) and
noticing that the numbers of types of $U_{XY}$ and $Q_{\tilde AB}$
     are polynomials of $n$, hence sub-exponential, we have:
\begin{eqnarray}
E(1(a^n \neq \widehat a^n(b^n, y^n) ))& \leq & \frac{1}{|\mathcal C|} \sum_{c\in \mathcal C}1^c(a^n \neq \widehat a^n(b^n, y^n) ) \nonumber\\
&\leq &\min\{1,2^{-n (I(S_X(Q_{A}); V_{Y|X})- H(Q_{A|B}
)-\epsilon^1_n)} \}\nonumber\\
& = & 2^{-n |I(S_X(Q_{A}); V_{Y|X})- H(Q_{A|B}
)-\epsilon^1_n|^+}\label{eqn.final_touch1}
\end{eqnarray}

Finally, we substitute~(\ref{eqn.final_touch1})
and~(\ref{eqn.lower_linearE1}) into~(\ref{eqn.def_lower}). Notice
that the number of types of $V_{Y|X}$ and $Q_{ AB}$ are polynomials
in $n$ and the usual method of types argument ( upper bounding the
probability of $P_{AB} (\rva^n,\rvb^n)\in Q_{AB})$ etc.), we have:
\begin{eqnarray}
p_e^n&=&\frac{1}{|\mathcal C|}\sum_{c\in \mathcal C}\Pr^c(\rva^n
\neq \widehat \rva^n(\rvb^n, \rvy^n))\nonumber\\
&\leq & \sum _{Q_{AB}, V_{Y|X}}2^{-n(D(Q_{AB}\|P_{AB})+ D(V_{Y|X}\|
W_{Y|X}|S_X(Q_A))+ |I(S_X(Q_{A}); V_{Y|X})- H(Q_{A|B}
)-\epsilon^1_n|^+ -\epsilon^2_n)}\nonumber\\
&\leq & \sum _{Q_{AB}, V_{Y|X}}2^{-n(D(Q_{AB}\|P_{AB})+ D(V_{Y|X}\|
W_{Y|X}|S_X(Q_A))+ |I(S_X(Q_{A}); V_{Y|X})- H(Q_{A|B} )|^+ -
\epsilon^1_n-\epsilon^2_n)}\nonumber\\
&\leq & 2^{-n(\min_{Q_{AB}, V_{Y|X}}\{D(Q_{AB}\|P_{AB})+ D(V_{Y|X}\|
W_{Y|X}|S_X(Q_A))+ |I(S_X(Q_{A}); V_{Y|X})- H(Q_{A|B} )|^+\} -
\epsilon^3_n)}\nonumber
\end{eqnarray}
where $\epsilon^i_n\rightarrow 0$ for $i=1,2,3$. Notice that $p_e^n$
is the average error probability of the codebook set $\cal C$, so
there exists at least a codebook $c$, such that the error
probability is no bigger than $p^n_e$.

Now we lower bound the achievable error exponent by

\begin{eqnarray}
&& \min_{Q_{AB}, V_{Y|X}}\{D(Q_{AB}\|P_{AB})+ D(V_{Y|X}\|
W_{Y|X}|S_X(Q_A))+ |I(S_X(Q_{A}); V_{Y|X})- H(Q_{A|B}
)|^+\}\nonumber\\
&=& \min_{Q_A}\min_{Q_{B|A}, V_{Y|X}}\{D(Q_{AB}\|P_{AB})+
D(V_{Y|X}\| W_{Y|X}|S_X(Q_A))+ |I(S_X(Q_{A}); V_{Y|X})- H(Q_{A|B}
)|^+\}\nonumber\\
&=& \min_{Q_A}\max_{S_X(Q_A)}\min_{Q_{B|A},
V_{Y|X}}\{D(Q_{AB}\|P_{AB})+ D(V_{Y|X}\| W_{Y|X}|S_X(Q_A))+
|I(S_X(Q_{A}); V_{Y|X})- H(Q_{A|B} )|^+\}\nonumber
\end{eqnarray}

The last equality is true because that the codeword composition
$S_X(Q_A)$ can be picked according to the source composition $Q_A$.
And by our code book selection we always pick the composition to
maximize the error exponent
$$\min_{Q_{B|A}, V_{Y|X}}\{D(Q_{AB}\|P_{AB})+
D(V_{Y|X}\| W_{Y|X}|S_X(Q_A))+ |I(S_X(Q_{A}); V_{Y|X})- H(Q_{A|B}
)|^+\}.$$

Here we slightly abuse the notations where $S_X(Q_A)$ is always the
optimal distribution to maximize the above exponent given $Q_A$.

The lower bound on $E_{}(P_{AB}, W_{Y|X})$ in
Theorem~(\ref{theorem:main}) is just proved. \hfill $\blacksquare$

\vspace{0.1in}
\subsubsection{Upper bound} \footnote{In this proof, $\epsilon^i_n>0$ and
$\epsilon^i_n\rightarrow 0$,  $i=1,2,3,4,4',5,6$ and $7$.} First we
fix the source composition $Q_A$, there are
$2^{n(H(Q_A)-\epsilon^1_n)}$ sequences in $\mathcal A^n$ with type
$Q_A$. When the encoder observes the source sequence $a^n$, it has
to send a code word $x^n(a^n)$ to the channel $W_{Y|X}$. There are
at most $(n+1)^{|X|}$ different types, so at least
$$\frac{2^{n(H(Q_A)-\epsilon^1_n)}}{(n+1)^{|X|}}=
2^{n(H(Q_A)-\epsilon^{2'}_n)}$$ of the codewords for $a^n\in Q_A$
have the same composition, we write this composition $S_X(Q_A)$, and
$$A_1=\{a^n\in Q_A: x^n(a^n)\in S_X(Q_A)\}, \mbox{  where } |A_1|=2^{n(H(Q_A)-\epsilon^2_n)}.$$

Now we fix the conditional type $Q_{B|A}$, so we have the marginal
$Q_B$ and the joint distribution $Q_{AB}$ determined by $Q_{A}$ and
$Q_{B|A}$.   Write
$$Q_{A|B}(b^n)=\{a^n: (a^n, b^n)\in Q_{AB}\} \mbox{ and } Q_{B|A}(a^n)=\{b^n: (a^n, b^n)\in Q_{AB}\}.$$
Obviously $|Q_B|=2^{n(H(Q_B)-\epsilon^3_nn)}$ and for all $b^n$:
$|Q_{A|B}(b^n)|=2^{n(H(Q_{A|B})-\epsilon^4_n)}$, for all $a^n$:
$|Q_{B|A}(a^n)|=2^{n(H(Q_{B|A})-\epsilon^{4'}_n)}$.

Let $B_1=\{b^n\in Q_B: |Q_{A|B}(b^n)\bigcap A_1|\geq
2^{n(H(Q_{A|B})-\epsilon^5_n)} \}$, where $\epsilon^5_n=
 \epsilon^2_n+\epsilon^{4'}_n+\frac{1}{n}$.  We   show next that the size of $B_1$ is of the
order $2^{nH(Q_B)}$.

Let $AB_1=\{(a^n, b^n): a^n\in A_1\mbox{ and } (a^n, b^n)\in
Q_{AB}\}$, we compute the size of $AB_1$ from two different ways.\\
First \begin{eqnarray}|AB_1|=|A_1||Q_{B|A}(a^n)|=
2^{n(H(Q_{AB})-\epsilon^2_n- \epsilon^{4'}_n)}
\label{eqn.boundingB1.0}.
\end{eqnarray}

Secondly
\begin{eqnarray}
 |AB_1|& =& |\{(a^n, b^n): b^n\in B_1, \  a^n\in A_1\mbox{ and } (a^n, b^n)\in
Q_{AB}\}\nonumber\\
&& \bigcup \{(a^n, b^n): b^n\in Q_B-B_1, \  a^n\in A_1\mbox{ and }
(a^n, b^n)\in Q_{AB}\}|
\label{eqn.boundingB1.1}\\
&\leq & |B_1||Q_{A|B}(b^n)|  + |Q_B-B_1|2^{n(H(Q_{A|B})-\epsilon^5_n)} \label{eqn.boundingB1.2}\\
&=&|B_1| 2^{n(H(Q_{A|B})-\epsilon^4_n)}  + (2^{n(H(Q_{B})-\epsilon^3_n)}-|B_1|)2^{n(H(Q_{A|B})-\epsilon^5_n)} \nonumber\\
&\leq &|B_1| 2^{nH(Q_{A|B})}  +
2^{nH(Q_{B})}2^{n(H(Q_{A|B})-\epsilon^5_n)} \label{eqn.boundingB1.4}
\end{eqnarray}
(\ref{eqn.boundingB1.1}) is by the definition of $AB_1$ and $B_1$,
(\ref{eqn.boundingB1.2}) is by the definition of $B_1$,
(\ref{eqn.boundingB1.4}) is true because all $\epsilon_n^i$'s are
positive.

Combining~(\ref{eqn.boundingB1.0}) and~(\ref{eqn.boundingB1.4}) and
use the fact that $\epsilon^5_n=
 \epsilon^2_n+\epsilon^{4'}_n+\frac{1}{n}$, we have:
\begin{eqnarray}
|B_1|2^{nH(Q_{A|B})} &\geq& 2^{n(H(Q_{AB})-\epsilon^2_n-
\epsilon^{4'}_n)}- 2^{nH(Q_{B})}2^{n(H(Q_{A|B})-\epsilon^5_n)}
\nonumber\\
&=& 2^{n(H(Q_{AB})-\epsilon^2_n- \epsilon^{4'}_n)}\times
\frac{1}{2}. \nonumber\end{eqnarray} Hence $|B_1|\geq
2^{n(H(Q_{B})-\epsilon^2_n- \epsilon^{4'}_n -\frac{1}{n})}=
2^{n(H(Q_{B}) - \epsilon^{5}_n)}$.

Now we consider the decoding error of the following events and show
that this error events gives us an upper bound on the error exponent
stated in this theorem:

\textbf{Source and side information pair $AB^*=\{(a^n, b^n): a^n\in
A_1, b^n\in B_1, (a^n, b^n)\in Q_{AB}\}$}.

First, for each $(a^n, b^n)\in AB^*$:
$$P_{AB}(a^n,b^n)=2^{-n(D(Q_{AB}\|P_{AB})+H(Q_{AB}))}.$$
Secondly, the size of $AB^*$ is lower bounded as follows from the
definition of $B_1$ and the lower bound on $|B_1|$:
\begin{eqnarray}
|AB^*|&\geq& |B_1|\times 2^{n(H(Q_{A|B})-\epsilon^5_n)}
\}\nonumber\\
&\geq&2^{n(H(Q_{B}) - \epsilon^{5}_n)}\times
2^{n(H(Q_{A|B})-\epsilon^5_n)}\nonumber\\
&\geq&2^{n(H(Q_{AB}) - 2\epsilon^{5}_n)}\label{eqn.size0fABstar}
\end{eqnarray}
So obviously the probability of $AB^*$ is
\begin{eqnarray}P_{AB}(AB^*)=
|AB^*|2^{-n(D(Q_{AB}\|P_{AB})+H(Q_{AB}))}\geq
2^{-n(D(Q_{AB}\|P_{AB})+2\epsilon^5_n)}.\label{eqn.probofABstar}\end{eqnarray}

Thirdly, if the side-information is $b^n\in B_1$ there are at least
$2^{n(H(Q_{A|B})-\epsilon^5_n)}$ many $a^n$'s such that
$(a^n,b^n)\in Q_{AB}$, that is, there are at least
$2^{n(H(Q_{A|B})-\epsilon^5_n)}$  many source sequences with the
same likelihood given the side-information $b^n$ (even there exists
a ``genie'' that tells the decoder that the joint distribution of
$(a^n, b^n)$ is $Q_{AB}$). Furthermore, the channel input codeword
$x^n(a^n)$ for these source sequences all have composition
$S_X(Q_A)$. Hence we have a channel coding problem with rate
$H(Q_{A|B})-\epsilon^5_n  $ and fixed input composition $S_X(Q_A)$.
This is the standard channel coding sphere packing bound studied
in~\cite{Gallager_sphere}.

So if $b^n\in B_1$, then \textbf{average} error probability for
$(a^n,b^n)\in AB^*$ is  at least:

\begin{eqnarray}
&&2^{-n(\min\limits_{V_{Y|X}: I(S_X(Q_A);V_{Y|X})<
H(Q_{A|B})-\epsilon^5_n}\{D(V_{Y|X}\|
W_{Y|X}|S_X(Q_A))\}+\epsilon^6_n)}\nonumber\\
&\geq & 2^{-n(\min\limits_{V_{Y|X}: I(S_X(Q_A);V_{Y|X})<
H(Q_{A|B})}\{D(V_{Y|X}\|
W_{Y|X}|S_X(Q_A))\}+\epsilon^7_n)},\label{eqn.average_error}
\end{eqnarray}
where $\epsilon^5_n$ and  $\epsilon^6_n$ goes to zero as $n$ goes to
infinity, hence $\epsilon^7_n\rightarrow 0$ because
$I(S_X(Q_A);V_{Y|X})$ is continuous in $V_{Y|X}$ and $D(V_{Y|X}\|
W_{Y|X}|S_X(Q_A))$ is convex in $V_{Y|X}$.

Finally we combine~(\ref{eqn.probofABstar})
and~(\ref{eqn.average_error}), and notice that the above analysis is
true for any(adversary) distribution of the source $Q_A$, and
any(optimal) channel codebook  composition $S_X(Q_A)$, and
any(adversary) $Q_{B|A}$  after $Q_A$ and $S_X(Q_A)$ are chosen,
 the error probability is lower bounded by:
\begin{eqnarray}
&&
2^{-n(\min\limits_{Q_A}\max\limits_{S_X(Q_A)}\min\limits_{Q_{B|A}}\{D(Q_{AB}\|P_{AB})+\min\limits_{V_{Y|X}:
I(S_X(Q_A);V_{Y|X})< H(Q_{A|B})}\{D(V_{Y|X}\|
W_{Y|X}|S_X(Q_A))\}\}+2\epsilon^5_n
+\epsilon^7_n)} \nonumber\\
 &=&   2^{-n(\min\limits_{Q_A}\max\limits_{S_X(Q_A)}\min\limits_{Q_{B|A},V_{Y|X}: I(S_X(Q_A);V_{Y|X})<
H(Q_{A|B})}\{D(Q_{AB}\|P_{AB})+D(V_{Y|X}\| W_{Y|X}|S_X(Q_A))
\}+2\epsilon^5_n +\epsilon^7_n)} \nonumber
\end{eqnarray}
Both $\epsilon^5_n $ and $\epsilon^7_n $ converges to zero as $n$
goes to infinity, the upper bound in Theorem~\ref{theorem:main} is
just proved.
 \hfill $\blacksquare$

\subsection{Proof of
Corollary~\ref{cor.digital}}\label{sec.appendix.digial} The proofs
for both lower bounds and uppers with the ``digital interface'' are
similar.

\subsubsection{Proof of~(\ref{eqn.digitallower}), the lower bound}

By introducing the auxiliary variable $R$ to separate the source
coding and channel coding error exponents and the definition of
error exponents, the following equalities should be obvious.

\begin{eqnarray}
&&E_{ }(P_{AB}, W_{Y|X})\nonumber\\
 &\geq&\min_{Q_A}\max_{S_X(Q_A)}\min_{Q_{B|A}, V_{Y|X}}
D(Q_{AB}\|P_{AB})+ D(V_{Y|X}\| W_{Y|X}|S_X(Q_A))
 +|I(S_X(Q_{A});
V_{Y|X})- H(Q_{A|B} )|^+ \nonumber\\
&=&\min_{Q_A}\max_{S_X(Q_A)}\min_{R}\nonumber\\
&&\{\min_{Q_{B|A}, V_{Y|X}: H(Q_{A|B} )=R} D(Q_{AB}\|P_{AB})+
D(V_{Y|X}\| W_{Y|X}|S_X(Q_A))
 +|I(S_X(Q_{A});
V_{Y|X})-R|^+\} \nonumber\\
&=&\min_{Q_A}\max_{S_X(Q_A)}\min_{R}\{ \min_{Q_{B|A}: H(Q_{A|B} )=R}
D(Q_{AB}\|P_{AB})  + E_r(R, S_X(Q_A),
W_{Y|X})\}\label{eqn.lower_Rinterp0}\\
&=&\min_{Q_A}\max_{S_X(Q_A)}\min_{R}\{  e'_U(R, P_{AB}, Q_A)+ E_r(R,
S_X(Q_A), W_{Y|X})\} \label{eqn.lower_Rinterp}
\end{eqnarray}
where $E_r(R, S_X(Q_A), W_{Y|X})$ is the standard random coding
error exponent for channel $W_{Y|X}$ at rate $R$ and input
distribution $S_X(Q_A)$, while $e'_U(R, P_{AB}, Q_A)$ is a peculiar
source coding with side-information error exponent for source
$P_{AB}$ at rate $R$, where the empirical source distribution is
fixed at $Q_A$. That is for $Q_A$
\begin{eqnarray}
e'_U(R, P_{AB}, Q_A)
 &\triangleq& \min_{Q_{B|A}:  H(Q_{A|B} )= R}
D(Q_{AB}\|P_{AB})\nonumber
\end{eqnarray}

(\ref{eqn.lower_Rinterp}) needs more examination. It is obvious that
\begin{eqnarray}
e'_U(R, P_{AB}, Q_A)
 &\geq& \min_{Q_{B|A}:  H(Q_{A|B} )\geq R}
D(Q_{AB}\|P_{AB})\triangleq e_U(R, P_{AB}, Q_A)\nonumber.
\end{eqnarray}
where $e_U(R, P_{AB}, Q_A)$ is defined in~(\ref{eqn.define_eu}). Now
(\ref{eqn.lower_Rinterp}) becomes
\begin{eqnarray}
 E_{ }(P_{AB}, W_{Y|X}) \geq \min_{Q_A}\max_{S_X(Q_A)}\min_{R}\{
e_U(R, P_{AB}, Q_A)+ E_r(R, S_X(Q_A), W_{Y|X})\}
\label{eqn.lowerfinal}.\hfill
\end{eqnarray}

\subsubsection{Proof of~(\ref{eqn.digitalupper}), the upper  bound}
Similar to the proof for~(\ref{eqn.digitallower}), we have the
following equalities:
\begin{eqnarray}
&&E_{ }(P_{AB}, W_{Y|X})\nonumber\\
&\leq&\min\limits_{Q_A}\max\limits_{S_X(Q_A)}\min\limits_{Q_{B|A},V_{Y|X}:
I(S_X(Q_A);V_{Y|X})< H(Q_{A|B})}\{D(Q_{AB}\|P_{AB})+D(V_{Y|X}\|
W_{Y|X}|S_X(Q_A)) \} \nonumber\\
&=&\min_{Q_A}\max_{S_X(Q_A)}\min_{R}\min\limits_{Q_{B|A},V_{Y|X}:
I(S_X(Q_A);V_{Y|X})<R < H(Q_{A|B})}\{D(Q_{AB}\|P_{AB})+D(V_{Y|X}\|
W_{Y|X}|S_X(Q_A)) \} \nonumber\\
&=&\min_{Q_A}\max_{S_X(Q_A)}\min_{R}\{\min\limits_{Q_{B|A}:
 H(Q_{A|B})>R}D(Q_{AB}\|P_{AB})+
\min\limits_{ V_{Y|X}: I(S_X(Q_A);V_{Y|X})<R}D(V_{Y|X}\|
W_{Y|X}|S_X(Q_A)) \} \nonumber\\
&=&\min_{Q_A}\max_{S_X(Q_A)}\min_{R}\{ \min_{Q_{B|A}: H(Q_{A|B} )>R}
D(Q_{AB}\|P_{AB})  + E_{sp}(R, S_X(Q_A),
W_{Y|X})\}\label{eqn.upper_Rinterp0}\\
&=&\min_{Q_A}\max_{S_X(Q_A)}\min_{R}\{  e_U(R, P_{AB}, Q_A)+
E_{sp}(R, S_X(Q_A), W_{Y|X})\} \label{eqn.upper_Rinterp}
\end{eqnarray}
where $E_{sp}(R, S_X(Q_A), W_{Y|X})$ is the standard sphere packing
bound defined in~(\ref{eqn.channel_sphere_ee}) and $e_U(R, P_{AB},
Q_A)$ is defined in~(\ref{eqn.define_eu}). \hfill $\square$

  \thispagestyle{empty} \pagestyle{empty}

\bibliographystyle{plain}

\bibliography{./SI_EE_Main}

\end{document}